\documentclass[twocolumn,dvipsnames,nofootinbib,prx,floatfix,superscriptaddress]{revtex4-1}

\usepackage{xcolor}
\usepackage{amsmath,amssymb}
\usepackage{appendix}
\usepackage{makecell}
\usepackage{relsize}
\usepackage{graphicx}
\usepackage{stmaryrd}
\usepackage{bm}
\usepackage{lipsum}
\usepackage{multirow}
\usepackage{physics}
\usepackage{etoolbox}

\setcellgapes{4pt}
\usepackage[caption=false, position=top]{subfig}

\usepackage[colorlinks,plainpages]{hyperref}
\hypersetup{urlcolor=blue,citecolor=blue, linkcolor=blue,hypertexnames=false}

\usepackage{xspace}
\usepackage{setspace}
\usepackage{bm}

\newcommand{\OmrefOoneOfouraULalzeroU}{8.6}
\newcommand{\OmrefOoneOfouraULaltwothU}{6.3}
\newcommand{\OmrefOoneOfouraULalthreeU}{1.0}
\newcommand{\OmrefOoneOfouraULmargU}{1.5}

\newcommand{\OmrefOoneOfouraULalzerolU}{2.8}
\newcommand{\OmrefOoneOfouraULaltwothlU}{2.0}
\newcommand{\OmrefOoneOfouraULalthreelU}{3.2}
\newcommand{\OmrefOoneOfouraULmarglU}{2.9}

\newcommand{\ptestOfouraAlzero}{-2.2}
\newcommand{\ptestOfouraAltwoth}{-1.6}
\newcommand{\ptestOfouraAlthree}{-0.1}
\newcommand{\sigOfouraAlzero}{5.8}
\newcommand{\sigOfouraAltwoth}{4.5}
\newcommand{\sigOfouraAlthree}{0.8}
\newcommand{\ptestOoneOfouraAlzero}{-1.3}
\newcommand{\ptestOoneOfouraAltwoth}{-1.2}
\newcommand{\ptestOoneOfouraAlthree}{-0.3}
\newcommand{\sigOoneOfouraAlzero}{4.7}
\newcommand{\sigOoneOfouraAltwoth}{3.5}
\newcommand{\sigOoneOfouraAlthree}{0.6}

\newcommand{\bbhMedian}{\ensuremath 0.8}
\newcommand{\bbhUpperError}{\ensuremath 1.3}
\newcommand{\bbhLowerError}{\ensuremath 0.5}
\newcommand{\bbhPower}{\ensuremath -9}
\newcommand{\bbhfinalMedian}{\ensuremath 0.8}
\newcommand{\bbhfinalUpperError}{\ensuremath 1.1}
\newcommand{\bbhfinalLowerError}{\ensuremath 0.5}
\newcommand{\bbhfinalPower}{\ensuremath -9}
\newcommand{\nsbhMedian}{\ensuremath 5.0}
\newcommand{\nsbhUpperError}{\ensuremath 5.6}
\newcommand{\nsbhLowerError}{\ensuremath 3.1}
\newcommand{\nsbhPower}{\ensuremath -11}
\newcommand{\nsbhrateMedian}{\ensuremath 30}
\newcommand{\nsbhrateUpperError}{\ensuremath 34}
\newcommand{\nsbhrateLowerError}{\ensuremath 19}

\newcommand{\bnsMedian}{\ensuremath 3.6}
\newcommand{\bnsUpperError}{\ensuremath 6.7}
\newcommand{\bnsLowerError}{\ensuremath 2.8}
\newcommand{\bnsPower}{\ensuremath -11}
\newcommand{\bnsrateMedian}{\ensuremath 61}
\newcommand{\bnsrateUpperError}{\ensuremath 113}
\newcommand{\bnsrateLowerError}{\ensuremath 48}

\newcommand{\cbcMedian}{\ensuremath 0.9}
\newcommand{\cbcUpperError}{\ensuremath 1.1}
\newcommand{\cbcLowerError}{\ensuremath 0.5}
\newcommand{\cbcPower}{\ensuremath -9}
\newcommand{\alphaMedianOThree}{\ensuremath 2.7}
\newcommand{\alphaUpperErrorOThree}{\ensuremath 2.2}
\newcommand{\alphaLowerErrorOThree}{\ensuremath 1.9}
\newcommand{\alphaMedianOFoura}{\ensuremath 3.2}
\newcommand{\alphaUpperErrorOFoura}{\ensuremath 1.1}
\newcommand{\alphaLowerErrorOFoura}{\ensuremath 1.0}

\newcommand{\betaMedianOFoura}{\ensuremath 5.1}
\newcommand{\betaUpperErrorOFoura}{\ensuremath 4.5}
\newcommand{\betaLowerErrorOFoura}{\ensuremath 4.6}

\newcommand{\zpeakMedianOFoura}{\ensuremath 2.6}
\newcommand{\zpeakUpperErrorOFoura}{\ensuremath 1.2}
\newcommand{\zpeakLowerErrorOFoura}{\ensuremath 1.3}

\usepackage{orcidlink}

\newcommand{\OGW}{\Omega_{\rm GW}}

\newcommand{\hOm}{{\hat\Omega}}

\makeatletter
\let\@LN\relax
\makeatother

\begin{document}

\title{Upper Limits on the Isotropic Gravitational-Wave Background from the first part of LIGO, Virgo, and KAGRA's fourth Observing Run}

\author{The LIGO Scientific Collaboration, The Virgo Collaboration, and The KAGRA Collaboration\\[0.5em]
A.~G.~Abac\,\orcidlink{0000-0003-4786-2698}}
\affiliation{Max Planck Institute for Gravitational Physics (Albert Einstein Institute), D-14476 Potsdam, Germany}
\author{I.~Abouelfettouh}
\affiliation{LIGO Hanford Observatory, Richland, WA 99352, USA}
\author{F.~Acernese}
\affiliation{Dipartimento di Farmacia, Universit\`a di Salerno, I-84084 Fisciano, Salerno, Italy}
\affiliation{INFN, Sezione di Napoli, I-80126 Napoli, Italy}
\author{K.~Ackley\,\orcidlink{0000-0002-8648-0767}}
\affiliation{University of Warwick, Coventry CV4 7AL, United Kingdom}
\author{C.~Adamcewicz\,\orcidlink{0000-0001-5525-6255}}
\affiliation{OzGrav, School of Physics \& Astronomy, Monash University, Clayton 3800, Victoria, Australia}
\author{S.~Adhicary\,\orcidlink{0009-0004-2101-5428}}
\affiliation{The Pennsylvania State University, University Park, PA 16802, USA}
\author{D.~Adhikari}
\affiliation{Max Planck Institute for Gravitational Physics (Albert Einstein Institute), D-30167 Hannover, Germany}
\affiliation{Leibniz Universit\"{a}t Hannover, D-30167 Hannover, Germany}
\author{N.~Adhikari\,\orcidlink{0000-0002-4559-8427}}
\affiliation{University of Wisconsin-Milwaukee, Milwaukee, WI 53201, USA}
\author{R.~X.~Adhikari\,\orcidlink{0000-0002-5731-5076}}
\affiliation{LIGO Laboratory, California Institute of Technology, Pasadena, CA 91125, USA}
\author{V.~K.~Adkins}
\affiliation{Louisiana State University, Baton Rouge, LA 70803, USA}
% ...

\collaboration{The LIGO Scientific Collaboration, the Virgo Collaboration, and the KAGRA Collaboration}

\begin{abstract}
We present results from the search for an isotropic gravitational-wave background using Advanced LIGO and Advanced Virgo data from O1 through O4a, the first part of the fourth observing run. %
This background is the accumulated signal from unresolved sources throughout cosmic history and encodes information about the merger history of compact binaries throughout the Universe, as well as exotic physics and potentially primordial processes from the early cosmos. Our cross-correlation analysis reveals no statistically significant background signal, enabling us to constrain several theoretical scenarios. %
For compact binary coalescences which approximately follow a 2/3 power-law spectrum, we constrain the fractional energy density to $\Omega_{\rm GW}(25 \, \mathrm{Hz}) \leq 2.0 \times 10^{-9}$ (95\% credibility), a factor of 1.7 improvement over previous results. %
Scale-invariant backgrounds are constrained to $\Omega_{\rm GW}(25 \, \mathrm{Hz}) \leq 2.8 \times 10^{-9}$, representing a 2.1$\times$ sensitivity gain. %
We also place new limits on alternative gravity theories predicting non-standard polarization modes and confirm that terrestrial magnetic noise sources remain below our detection threshold. %
Combining these spectral limits with population models for GWTC-4, the latest gravitational-wave event catalog, we find our constraints remain above predicted merger backgrounds but are approaching detectability. The joint analysis combining the background limits shown here with the GWTC-4 catalog enables improved inference of the binary black hole merger rate evolution across cosmic time. %
Furthermore, employing GWTC-4 inference results and {standard modeling choices}, we estimate that the total background arising from compact binary coalescences is $\Omega_\mathrm{CBC}(25 \mathrm{Hz}) = {\cbcMedian^{+\cbcUpperError}_{-\cbcLowerError} \times 10^{\cbcPower}} $ at 90\% confidence, where the largest contribution is due to binary black holes only, $\Omega_\mathrm{BBH}(25\,\mathrm{Hz}) = \bbhfinalMedian^{+\bbhfinalUpperError}_{-\bbhfinalLowerError}\times 10^{\bbhfinalPower}$.

\end{abstract}

\maketitle

\section{Introduction}
\label{sec:intro}
The superposition of individually unresolved gravitational-wave (GW) signals accumulated throughout cosmic history gives rise to a gravitational-wave background (GWB)~\cite{Detection_methods_Joe_Neil, PhysRevD.92.042001} which produces a persistent incoherent signal in GW detectors. %
The GWB is commonly referred to as ``stochastic" as it can be characterized statistically as a mean zero stochastic field~\cite{Detection_methods_Joe_Neil}. %

GWB sources can be categorized as cosmological or astrophysical based on their origin. The former provides a unique probe of the early Universe, as GWs propagate directly from the inflationary epoch, while the latter results from an ensemble of unresolved astrophysical events, offering insights into source populations at high redshift. %
Potential contributors to the GWB include distant unresolved compact binary mergers~\cite{PhysRevD.84.124037,Zhu_CBC,PhysRevD.84.084004, PhysRevD.85.104024,PhysRevD.84.124037, Zhu_CBC_2}, core-collapse supernovae~\cite{PhysRevD.73.104024, PhysRevD.72.084001, PhysRevD.84.124037, Zhu_CBC_2}, rotating neutron stars~\cite{Zhu:2010af, PhysRevD.72.084001, PhysRevD.73.104024, 8201910, PhysRevD.87.063004, PhysRevD.86.104007, Zhu:2011pt}, stellar core collapse~\cite{Tania_CC, Tania_CC_2} and superradiance of axion clouds around black holes~\cite{Brito:2015oca,Takahashi:2024fyq, PhysRevLett.117.061101, PhysRevD.83.044026, PhysRevLett.119.131101}. Some of the cosmological sources are cosmic strings~\cite{Kibble_1976, Hindmarsh:1994re, Vachaspati:2015cma,Schmitz_CS, PhysRevD.97.102002, PhysRevLett.126.241102, Auclair:2019wcv, PhysRevLett.85.3761, PhysRevD.71.063510,Siemens:2007,Olmez:2010,Vachaspati:1985,Blanco-Pillado:2017,Kawasaki:2010}, first-order phase transitions in the early Universe~\cite{PhysRevD.30.272, Hogan:1986qda, Mazumdar:2018dfl,Hindmarsh:2020hop, PhysRevLett.65.3080, PhysRevD.45.4514, PhysRevD.49.2837, PhysRevLett.126.151301, PhysRevD.107.023511}, primordial black holes~\cite{1975ApJ...201....1C,Carr:2020gox,Matarrese:1997ay, Baumann:2007zm,Kohri:2018awv,PhysRevLett.128.051301, SIGWB_Ryoto, zphk-3ld9}, a stiff equation of state in the early Universe~\cite{Giovannini:1998bp,Peebles:1998qn,Giovannini:1999bh,Figueroa:2018twl,Figueroa:2019paj,PhysRevD.110.103503}, and gravitational waves generated during inflation~\cite{Barnaby:2011qe,Barnaby:2011vw,Maleknejad:2016qjz,Thorne:2017jft,Bartolo:2016ami,Starobinsky:1979ty, PhysRevD.110.084063, Mukhanov:1981xt, Grishchuk:1974ny, RUBAKOV1982189, ABBOTT1984541}.

The Laser Interferometer Gravitational-wave Observatory (LIGO)~\cite{Aasi_2015}, Virgo~\cite{Acernese_2015}, and KAGRA~\cite{KAGRA:2020tym} collaborations (LVK collaborations~\cite{2018LRR....21....3A}) have previously placed upper limits on both isotropic~\cite{PhysRevD.104.022004} and anisotropic~\cite{PhysRevD.104.022005} GWBs using data from the first three observing runs collected by the LIGO Hanford, LIGO Livingston, and Virgo interferometers. These searches relied on cross-correlating data between detector pairs~\cite{Detection_methods_Joe_Neil}. %
The implications of the upper limits placed on the GWB energy density spectrum with LVK data have been far-reaching, setting competitive bounds on early Universe physics~\cite{PhysRevD.97.102002, PhysRevLett.126.151301, PhysRevLett.128.051301, SIGWB_Ryoto, PhysRevLett.126.241102, PhysRevD.110.103503, PhysRevD.110.084063, PhysRevD.107.023511}, and expanding our knowledge of the astrophysical binary population properties complementary to what is obtained from individual GW sources~\cite{Callister:2020,PhysRevD.104.022004, Turbang:2023tjk, Lalleman:2025xcs}. 

The most compelling astrophysical sources of a GWB in our detector range are compact binary coalescences (CBCs). 
The CBC background is expected to be an approximately isotropic~\cite{Cusin:2018rsq, Jenkins:2018uac, Capurri:2021zli} spectral power-law signal, peaking around a few hundred Hz (depending on the features of the CBC population~\cite{Regimbau:2011,Capurri:2021zli,Renzini:2024,Ebersold:2025}). %
The signal is not expected to be Gaussian due to the intermittent nature of the background, in the frequency range probed~\cite{Drasco:2002yd}. %
In this work, we update our models to estimate the expected contributions from binary black holes (BBHs), binary neutron stars (BNSs) and neutron star-black hole (NSBH) coalescences, incorporating the most recent observational data from O4a~\cite{GWTC-4-Introduction,GWTC-4-Methods,GWTC-4-Results,GWTC-4-Population}. For the BBH population in particular, we adopt the methodology of~\cite{Callister:2020} to constrain the redshift-dependent merger rate by combining individually resolved events with upper limits from the cross-correlation analysis presented in this paper. This approach allows us to construct a full posterior distribution for the resulting energy density spectrum, accounting for current uncertainties in the mass, spin and redshift distributions of the BBH population. Additionally, by combining the contributions from all classes of CBCs and comparing the resulting GWB spectrum with the projected sensitivity curves of the upgraded Advanced LIGO detectors (A+)~\cite{aplus_design} and Virgo detector (V+)~\cite{O5VirgoNoiseASDCurve}, we demonstrate that one year of time-coincident data between detectors in the network at design sensitivity would enable us to substantially probe and potentially detect the expected signal arising from the CBC population. %

In this work we follow the same methodology as previous observing runs, cross-correlating data from LIGO Hanford and Livingston detectors during the first portion of the fourth observing run, hereinafter referred to as O4a. During this first part of the run, Virgo was not in science mode, hence why it is not included in our analysis. In this paper, we focus exclusively on the isotropic GWB, while complementary results from an anisotropic search are presented in a companion paper~\cite{Anisotropic}. %
The cross-correlation search method employed is optimal for a continuous Gaussian signal, given appropriate caveats~\cite{Detection_methods_Joe_Neil, Sufficient_statistic_Andrew}, while it is sub-optimal but still sensitive to an intermittent signal~\cite{Lawrence:2023buo}. Several alternative search methods are currently being explored tailored to intermittent signals~\cite{smithOptimalSearchAstrophysical2018,Buscicchio:2022raf,Lawrence:2023buo,Sah:2023bgr}.

As we do not detect a GWB, we set upper limits on its amplitude. We include inference on a set of possible GWB models, including an astrophysical signal sourced by CBCs and alternative GWB polarization modes~\cite{PhysRevD.79.082002} arising in theories of gravity beyond General Relativity (GR). As in O3, loud glitches are excluded from the analysis~\cite{gatingDocument, y_stochasticGatingDocument}. Additionally, as correlated magnetic noise is a potential limiting factor in stochastic background searches especially as detector sensitivity improves, we construct a magnetic noise budget, and find it lies well below the current GWB sensitivity. %
Alongside this paper, the LVK collaboration is also releasing a companion paper exploring the cosmological implications of our results~\cite{CosmoImplications}, focusing on the constraints these place on potential early-Universe sources of the GWB. %

This paper is organized as follows. In Section~\ref{sec:methods}, we outline the analysis methods employed throughout. %
Section~\ref{sec:dq} details the procedures implemented to remove identifiable sources of detector artifacts that can bias the results and implications of the search. Section~\ref{sec:results} presents the results of our analysis. Finally, in Section~\ref{sec:astro}, we discuss the astrophysical implications of our findings and provide constraints on the BBH merger rate.

\section{Methods}
\label{sec:methods}
The isotropic stochastic analysis presented here targets the spectral properties of the GWB, modeling the signal as isotropic, unpolarized, stationary, and Gaussian in the limit of long observing time~\cite{Detection_methods_Joe_Neil, renzini_review}. %
The signal is also assumed to be weak with respect to the detector sensitivity~\cite{Sufficient_statistic_Andrew}. 

The GWB spectrum may be expressed in terms of the fractional GW energy density spectrum $\Omega_{\rm GW}(f)$, defined as
\begin{equation}
    \Omega_{\rm GW}(f) = \frac{1}{\rho_\mathrm{c}} \frac{\text{d}\rho_{\rm GW} (f)}{\text{d}\ln f}\,,
\label{eq:Omegarho}
\end{equation}
where $\text{d}\rho_{\rm GW}$ is the GW energy density in the frequency band $[f, f+\text{d}f]$, and $\rho_{\rm c}= 3 H_0^2 c^2/(8\pi G)$ is the critical energy density in the Universe, with $c$ the speed of light and $G$ Newton’s constant. Integrating $\Omega_{\rm GW}(f)$ over $\text{d}\ln f$ returns the total dimensionless GW energy density. %
In what follows, we adopt the Planck 2015 cosmological parameters, fixing the Hubble constant to $H_0 = 67.9 \, \mathrm{km} \, \mathrm{s}^{-1} \, \mathrm{Mpc}^{-1}$ and the present-day matter density parameter to $\Omega_\mathrm{m,0} = 0.3065$~\cite{Planck:2015}. %

The analysis presented is carried out with the {\tt pygwb} analysis package~\cite{Renzini:2023qtj,Renzini2024}. %
In the following we summarize the methodological approach and specific choices made in the analysis of the O4a dataset. %
Methods employed here do not vary significantly from those of previous observing runs O1~\cite{PhysRevLett.118.121101}, O2~\cite{PhysRevD.100.061101}, and O3~\cite{PhysRevD.104.022004}, so that the resulting spectra may be ultimately combined into a single result. %
Changes with respect to previous analyses are highlighted in the text. %


\subsection{Cross-correlation spectra}

\label{sec:methods:cross-corr}

Following~\cite{PhysRevD.59.102001}, we employ cross-correlation spectra calculated from pairs of independent, time-coincident detector strain datastreams to construct an estimator for $\OGW(f)$:
\begin{equation}
    \hOm_{0}(f) = \frac{\text{Re}[C_{IJ}(f)]}{\gamma_{IJ}(f) S_0(f)}\,,
\label{eq:hom_gw}
\end{equation}
where $C_{IJ}$ is the cross-spectral density (CSD) calculated from the Fourier-transformed strain data $\tilde{s}(f)$ of duration $T$ from two non-coincident detectors $I$ and $J$: 
\begin{equation}
    C_{IJ}(f) = \frac{2}{T} \tilde{s}_I^*(f) \tilde{s}_J(f). 
\end{equation}

The geometric quantity $\gamma_{IJ}(f)$ is referred to as the overlap reduction function (ORF)~\cite{Finn:2008vh, Detection_methods_Joe_Neil}, which quantifies the frequency-dependent decoherence that affects the cross-correlation of data from two geographically separated interferometers. This reduction depends on the individual detector responses as well as their relative separation and orientation~\cite{Detection_methods_Joe_Neil}. The function $S_0(f)$ converts units of GW strain power into fractional energy density, whilst encoding also a frequency weighting specific to a chosen spectral model $F_0(f)$: 
\begin{align}
    S_0(f) = \frac{3 H_0^2F_0^2(f)}{10\pi^2}\frac{1}{f^3}\,.
\end{align}
As discussed in~\cite{Sufficient_statistic_Andrew}, in the case of noise-dominated data streams the cross-correlation statistic described above is a quasi-optimal~\footnote{Sub-optimality is due to corrections to the variance Eq.~\eqref{eq:variance} of the same order of magnitude of the signal, which become negligible when the signal is weak compared to the detector noise.} estimator for the background spectrum. %
 
In line with previous analyses~\cite{PhysRevLett.118.121101, PhysRevD.100.061101,PhysRevD.104.022004}, here we assume a power-law spectral model expressed as an amplitude $\Omega_{\alpha}$ with respect to the fixed reference frequency $f_{\rm ref}=25\,{\rm Hz}$ scaling with a power-law index $\alpha$,
\begin{equation}
    \Omega_{\rm GW}(f) = \Omega_{\alpha} F_{\alpha}(f)\,,\qquad
    F_{\alpha}(f) = \qty(\frac{f}{25\,{\rm Hz}})^\alpha\,.
    \label{eq:plaw_model}
\end{equation}
In Eq.~\eqref{eq:hom_gw} specifically, the chosen spectral model is fixed to a power law with spectral index $\alpha=0$, as indicated by the subscript of $S_0$ and that of the estimator in Eq.~\eqref{eq:hom_gw}. %
The corresponding variance estimate is~\cite{Detection_methods_Joe_Neil}
\begin{equation}
    \sigma_0^2(f) = \frac{1}{2 T \Delta f}
        \frac{P_I(f) P_J(f)}{\gamma_{IJ}^2(f) S_0^2(f)}\,,
        \label{eq:variance}
\end{equation}
where $P_I(f)$ is the one-sided power spectral density for detector $I$, and $\Delta f$  is the frequency resolution employed in the analysis. %

In practice, due to noise non-stationarities and computational efficiency the $\hOm_0(f)$ spectrum is estimated from a weighted average over a large number of data segments, where for each time segment $i$ there is a corresponding CSD estimate $C_{IJ,\,i}$ and PSD estimates $P_{I,\,i}$, $P_{J,\,i}$.  %
Further details about this cross-correlation computation may be found in the {\tt pygwb} publication~\cite{Renzini:2023qtj}.

\subsection{Frequency estimator}
\label{sec:methods:estimator}
We construct a narrow-band estimator $\hOm_0(f)$ with associated variance $\sigma_0^2(f)$: 
\begin{align}
    \hOm_0(f) &= \frac{\sum_i \qty[ \hOm_0(f; t_i) \sigma^{-2}_0 (f; t_i)]}{\sum_i \sigma^{-2}_{0} (f; t_i)}\,, \label{eq:hOmGW}\\
    \sigma_0(f) &= \left[\sum_i \sigma^{-2}_0 (f; t_i)\right]^{-1/2}\,,
    \label{eq:sigmaGW}
\end{align}
where $i$ labels data from $t_i$ to $t_i+\Delta t$. %
In practice, the cross-spectral densities are calculated from the data with a 50\% overlap to make up for the impact of Hann-windowing each segment, hence each independent time estimate is achieved by handling even and odd segments separately as described in~\cite{Renzini:2023qtj}. %

This calculation is performed employing the $S_0(f)$ weighting function. %
However, note that it can be performed by choosing any other power-law weighting $\alpha$, and it is possible to re-weight from one index, $\alpha_1$, to another, $\alpha_2$, as
\begin{equation}
    \hOm_{\alpha_1} (f) = \hOm_{\alpha_2} (f) \frac{F_{\alpha_1}(f)}{F_{\alpha_2}(f)}\,.
\end{equation}

Broadband estimates for a spectral index $\alpha$ and at reference frequency $25\,{\rm Hz}$ are obtained by optimally weighting the narrowband estimator, as described in~\cite{PhysRevD.104.022004,Renzini:2023qtj}:
\begin{align}
    \hOm_{\alpha} |_{25\,{\rm Hz}} &= \frac{\sum_f \hOm_0(f)F_{\alpha}(f)\sigma_0^{-2}(f)}{\sum_f F^2_{\alpha}(f)\sigma_0^{-2}(f)} \,,\\
    \sigma_{\alpha} |_{25\,{\rm Hz}}&= \left[  \sum_f F_{\alpha}(f) \sigma_0^{-2} (f) \right]^{-1/2}\,.
    \label{eq:broadband_est}
\end{align}

\subsection{Parameter estimation}
\label{sec:methods:pe}
Parameter estimation is performed here using a hybrid approach that combines frequentist and Bayesian analysis techniques~\cite{PhysRevLett.109.171102}. Specifically, the frequentist estimators from the previous sections, $\hOm_0(f)$ and $\sigma_0(f)$, combined over the entire observing period, are used as input data to calculate posterior probability distributions of GWB model parameters. These in turn are used to compute upper bounds on the spectral amplitude of the GWB. %
In~\cite{Sufficient_statistic_Andrew} the authors show that in the weak-signal approximation this hybrid approach does not lose information compared to a fully Bayesian search, which produces posterior distributions from the full time series data rather than from the time-averaged frequency estimators. %

We assume that, in each frequency bin, the un-weighted spectral estimate $\hOm_0$ is Gaussian-distributed with variance $\sigma^2_{0}$~\cite{PhysRevD.104.022004}:
\begin{equation}
 p(\hOm_0|\Theta)
\propto
\exp
\left[
  -\frac{1}{2}\sum_f \frac{(\hOm_0(f)-\OGW(f;\Theta))^2}{\sigma_0^2(f)}
\right] \,,
\label{eq: SGWB _likelihood}
\end{equation}
where $\OGW(f;\Theta)$ is an assumed model for the GWB signal, parametrized via a set of model parameters $\Theta$. %
Posteriors for the model parameters $\Theta$ are obtained by employing the {\tt pygwb} library~\cite{Renzini:2023qtj} and the {\tt Bilby} library~\cite{Ashton:2018jfp} for detector modeling and likelihood evaluation. %
As in previous work, e.g.~\cite{PhysRevD.104.022004}, this analysis is performed using data from multiple detector pairs, $I, J$ (i.e., \emph{baselines}). %
Data from different baselines are assumed to be independent, such that the combined inference assumes likelihood~\eqref{eq: SGWB _likelihood} holds for data from each individual baseline, $\hOm_0^{IJ}(f),\,\sigma_{0}^{IJ}(f)$, and the total likelihood is the result of the product of the likelihood calculated for each baseline. %

We consider different models for the GWB:
\begin{itemize}
    \item Noise: $\Omega_{\rm GW}(f;\Theta) =  0$. We implicitly include uncorrelated Gaussian noise as part of every model that follows in the prior choices for the model parameters. This noise model is not strictly correct as there are potential sources of correlated noise, such as Schumann resonances~\cite{PhysRevD.107.022004}. However, these effects may be neglected in the present search due to detector sensitivity, as shown in Sec.~\ref{sec:dq:magnetic}.
    \item Power law: $\Omega_{\rm GW}(f;\Theta) = \Omega_{\rm ref}F_\alpha(f)$, as per Eq.~\eqref{eq:plaw_model}. %
    In this analysis, the two parameters $\Theta = \{\Omega_{\rm ref},\alpha \}$ are inferred independently, and hence we choose this notation to avoid confusion with the cross-correlation estimators described above. %
    The reference frequency is set to $f_{\rm ref}=25\,{\rm Hz}$ as in past analyses~\cite{PhysRevD.104.022004} due to the location of peak stochastic sensitivity, for lower values of $\alpha$. Note that the precise sensitivity depends on the spectral index $\alpha$. %
    \item Scalar-vector-tensor (SVT) power law: this model assumes that the signal may include vector and scalar polarizations as well as the standard tensor polarization predicted by General Relativity. This background is modeled as a superposition of individual contributions, labeled $\rm (p)=\{T,V,S\}$ where each component is individually modeled by a power law~\cite{Callister:2017ocg}: 
    \begin{equation}
    \Omega_{\rm GW}(f;\Theta) = \sum_{\rm (p)} \beta^{\rm (p)}(f)\Omega_{\rm ref}^{\rm (p)}F_{\alpha_{\rm p}}(f)\,,
    \end{equation}
    where $\beta^{\rm (p)}(f) \equiv \gamma_{IJ}^{\rm (p)}(f)/\gamma_{IJ}(f)$ is the ratio between the ORF for polarization (p) and the standard (tensor) polarization~\cite{PhysRevD.79.082002}.
    \item Magnetic: describes correlations between pairs of detectors induced by large-scale coherent magnetic fields that can mimic a GWB signal. These are modeled in terms of magnetometer correlations and a transfer function between the local magnetic field and the strain channel of the detectors. Further discussion on this model is given in section \ref{sec:results:jointMagGWB}.
    \item Compact binary coalescences (CBC): the predominant contribution to the GWB is expected to be the superposition of all merging compact binaries in the observable Universe~\cite{Regimbau:2011}.
    The total GW fractional energy density due to compact binary mergers, $\Omega_{\rm CBC}$, can be parametrized as a function of the mass distribution of compact binaries and their merger rate as a function of redshift. This model will be discussed in Section \ref{sec:astro}.
    Assuming $\Omega_{\rm CBC}$ is the only contributor to the stochastic signal in the analyzed frequency band, we can infer mass and merger rate population parameters.
\end{itemize}

To test these different models against the data, we calculate Bayes factors for each model. Unless otherwise specified, we compute the Bayes factor between the hypothesis of a specific signal model with respect to the noise model, $\Omega_{\rm GW}(f;\Theta) \equiv 0$.

\section{Data Quality}
\label{sec:dq}

This section describes the data quality and handling specifics for the strain data collected during the first portion of the fourth observing run (O4a) by the LIGO Hanford and Livingston detectors. %
Note that results presented in the following sections include the combination of this data with previous datasets O1, O2, O3, whose relevant instrument performances and data quality features are described in~\cite{Abbott:2016xvh,Martynov:2017ufx,Nuttall:2015dqa,LIGOScientific:2017tza},~\cite{LIGO:2021ppb,LIGOScientific:2021kro,LIGOScientific:2018kdd},~\cite{PhysRevD.102.062003,PhysRevLett.123.231107,Nguyen_2021,Soni_2021} respectively.  The O3 analysis also includes data from Virgo~\cite{ENV_noise,PhysRevLett.131.041403,PhysRevLett.123.231108,PhysRevLett.125.131101,Virgo:2022ypn,Virgo:2022ysc}. %
For further details on the performance of the LIGO detectors and the quality of their data during O4a~\cite{Capote:2024rmo,membersoftheLIGOScientific:2024elc}, see also~\cite{Detchar_paper_O4a}.

\subsection{Data}
\label{sec:dq:data}
The O4a observing period ran from 15:00 UTC on May 24, 2023 to 16:00 UTC on January 16, 2024. %
The total coincident time for the Hanford-Livingston (HL) baseline in this period was 126.57 days, before applying data quality cuts. %
The strain data used in this analysis were calibrated strain data with narrow-band features from photon calibrator and actuation injections and power
mains noise,  subtracted \cite{Viets2021Narrowband}, followed by the application of non-stationary noise subtraction \cite{Vajente_2020}. 
The correlated magnetic noise was monitored using low-noise LEMI120 magnetometers located at each site \cite{lemi}. 
The same data processing was carried out for both strain and magnetometer data to allow a comparison between the magnetic and GW analyses.

Both strain and magnetic datastreams are initially downsampled from the original sample rate of 16384 Hz to 4096 Hz. %
This downsampling is used because the GW sources targeted in this search populate frequencies below $2$ kHz. %
The data are high-pass filtered using a 16th-order Butterworth filter with a knee frequency of 11 Hz, which is constructed using second-order sections. %
As in analyses described in~\cite{PhysRevD.104.022004, Renzini:2023qtj}, each data stream is divided into 50$\%$ overlapping, Hann-windowed 192-second segments, coarse-grained to a frequency resolution of 1/32 Hz. %
The analyzed frequency range spans $20 - 1726$ Hz; the minimum frequency is chosen based on the detector sensitivities~\cite{Detchar_paper_O4a}, while the maximum frequency is chosen to be sufficiently below the Nyquist frequency to avoid aliasing effects. %


In the following text, tables, and figures, we label the LIGO Hanford detector as LHO and the LIGO Livingston as LLO.

\subsection{Time domain cuts: vetoed times}
\label{sec:dq:cuts}

Data quality for analysis purposes was assessed jointly with the Detector Characterization groups of LIGO, Virgo, and KAGRA. %
Data quality flags are employed to define observing time periods to be vetoed from analysis. %
Most significant to this search are the ``Category 1'' vetoes. %
These vetoes correspond to times when the detector was known to be operating outside of its nominal condition. %
For more details on the O4a data quality and the specific time-domain cuts made for stochastic data quality purposes, please refer to~\cite{Detchar_paper_O4a}, Sec. 4.3. %
After applying vetoes, the remaining viable time-coincident data is {126.47} days. %
The relevant segment information used to exclude vetoed data from this analysis will be available on GWOSC~\cite{LIGOScientific:2025snk}.

\subsection{Time domain cuts: gating} 
\label{sec:dq:gating}

In the third LVK observing run (O3), a large population of loud glitches~\cite{PhysRevD.104.022004} led to very frequent flagging of data segments by non-stationarity checks (see Sec.~\ref{sec:dq:dsc}). %
To mitigate this effect, time-domain gating of the data was introduced~\cite{GatingO3_2, GatingO3}: loud glitches are removed from the data stream by multiplying the data with an inverse Planck-taper window with parameters tailored to the specific non-stationary feature (e.g., duration)~\cite{McKechan:2010kp}. 

In O4a, data are preprocessed with an auto-gating procedure as described in~\cite{Renzini2024}. %
The parameters chosen for the auto-gating were optimized for this specific search, minimizing the amount of data lost whilst retaining stationarity. %
A total of $0.093\%$ of analyzed Hanford data and $0.068\%$ of Livingston data were gated; more details can be found in Sec. 4.3.2. of~\cite{Detchar_paper_O4a}.  %

\subsection{Time domain cuts: stationarity checks}
\label{sec:dq:dsc}
As in previous runs, we apply a non-stationarity cut (often referred to as the $\Delta\sigma$ cut~\cite{Renzini:2023qtj}) by removing times where the square root of the variance in Eq.~\eqref{eq:variance} is found to vary by more than $20 \%$ between neighbouring segments. %
We cut the union of segments that do not satisfy the stationarity test assuming a set of different spectral weights: $\alpha= \{ -5, 0, 3, 5\}$, implemented as in Eq.~\eqref{eq:hom_gw}. %
As each power law accumulates SNR in a different frequency band, these may flag different segments depending on their specific spectral narrowband/broadband features. %
This cut additionally removes 8.1$\%$ of the analysed segments, leaving 108.41 days of viable data.

\subsection{Frequency domain cuts: narrowband features}

\label{sec: notch_list}
In addition to broadband features that determine the exclusion of entire data segments from the analysis, also narrowband deviations from Gaussianity are investigated, and data cuts are applied to exclude problematic frequency bins. %
These frequencies are determined by monitoring the coherence between the GW channels at different sites, and between the GW channel and auxiliary channels at the same site. %
Bins that show evidence of instrumental noise or that manifest strong non-Gaussian behavior are excluded from the analysis: for more details, see~\cite{Detchar_paper_O4a}, Sec. 2.1. %
The list of problematic frequencies, i.e., the {\it notch list}, is determined at the end of the observing run and then applied identically throughout the entire data analysis process (with a small exception in parameter estimation, as discussed in Sec.~\ref{sec:results:upperLimits}). %
This list is included in the appended data release.

The notches include the following components: 
\begin{itemize}
    \item Calibration lines: these are stationary lines injected into each detector's differential arm feedback loop so as to continually monitor the detector response and the corresponding calibration amplitude factor. These include the stationary calibration lines injected throughout O4a~\cite{Karki:2016pht}, and one calibration line at $24.5$ Hz that was turned on between July 25, 2023, and August 9, 2023. 
    
    \item Pulsar injections: each injection corresponds to a simulation of a quasi-monochromatic, persistent GW signal as expected from a rapidly rotating, isolated neutron star. These were injected throughout the entire O4a run at both LIGO sites and emulate how canonical continuous-wave (CW) signals would appear in the detectors~\cite{PhysRevD.95.062002}. 

    \item Quadruple suspension violin modes: resonances of the detector mirror suspension fibers~\cite{Aasi_2015}.

    \item Powerline harmonics: integer harmonics of the 60 Hz power mains present at both detectors.
    \item Other artifacts: DuoTone signals (960Hz, 961Hz) and their 1 Hz side band present in both LIGO detectors in O4a from the timing system, used to synchronize data collection across the global detector network and in each interferometer~\cite{Sullivan:2023cqg}.
\end{itemize}

Overall, the notch list cuts 8.4\% of the spectrum; however, note that in the frequency range $[20,300]$ Hz where the detectors are most sensitive, only 2.7\% of the spectrum is cut. %

\subsection{Magnetic noise budget}
\label{sec:dq:magnetic}
Correlated magnetic noise present across multiple detectors is a probable hindrance for stochastic analyses~\cite{Thrane_2013, Thrane_2014, Coughlin_2018}.
Examples of such correlated magnetic noise are Schumann resonances~\cite{1952ZNatA...7..149S, 1952ZNatA...7..250S}, standing waves in the cavity between the Earth's surface and the ionosphere that may be excited by lightning strikes; and at higher frequencies, $f > 100$ Hz, correlated fields from lightning strikes.
Magnetic fields of any source couple to the detector via electric cables in the detector buildings or couple to the actuation magnets controlling the detector test mass mirrors~\cite{ENV_noise}.

Magnetic fields are monitored with sensitive probes placed outside the detector buildings to produce a magnetic ``noise budget", quantifying the strength of correlated magnetic noise present across detectors, which is then compared to the sensitivity of the detector network to a stochastic signal~\cite{Janssens:2022tdj}.
If this budget becomes large enough, correlated magnetic fields can contribute to correlated noise which may mimic a signal in the stochastic analysis presented here.

We compute the magnetic correlated noise budget via
\begin{equation}
    \hat{\Omega}_{\text {mag, } I J}(f)=\frac{2}{T} \frac{\left|T_I(f)\right|\left|T_J(f)\right| \sqrt{\sum_{ab} |\tilde{m}_{I,a}^{\star}(f) \tilde{m}_{J,b}(f)|^2}}{\gamma_{I J}(f) S_0(f)},
    \label{eq:mag_coup}
\end{equation}
where $T_I(f)$ and $T_J(f)$ are the magnetic coupling functions of detectors $I$ and $J$, and $\tilde{m}_{I, a}(f)$ is the Fourier-transformed magnetic data from the outside probes of detector $I$ pointing in the direction $a$, which can be either along the X or Y arms of a detector, and its sum runs over the four possible combinations of $a$ and $b$. %
Magnetic coupling functions consist of the product of two terms: an outside-to-inside coupling, $T_{\rm OTI}$, and an inside-to-strain coupling, $T_{\rm ITS}$. %
The $T_{\rm OTI}$ term is given by the ratio of the strength of a magnetic source measured outside the detector buildings to the strength of the same source measured inside the buildings, and reduces to an overall constant averaged over several independent measurements, while $T_{\rm ITS}$ is estimated via hardware injections as explained below. %
Uncertainties on these measurements include the uncertainty on the outside-to-inside coupling as measured in O3 at LLO, and, to be conservative, an intrinsic uncertainty factor of 2 as the magnetometers in the building (used for the measurements) are not located at the exact location where the magnetic field couples to the strain. 

\begin{figure}
    \centering
    \includegraphics[width=0.9\linewidth]{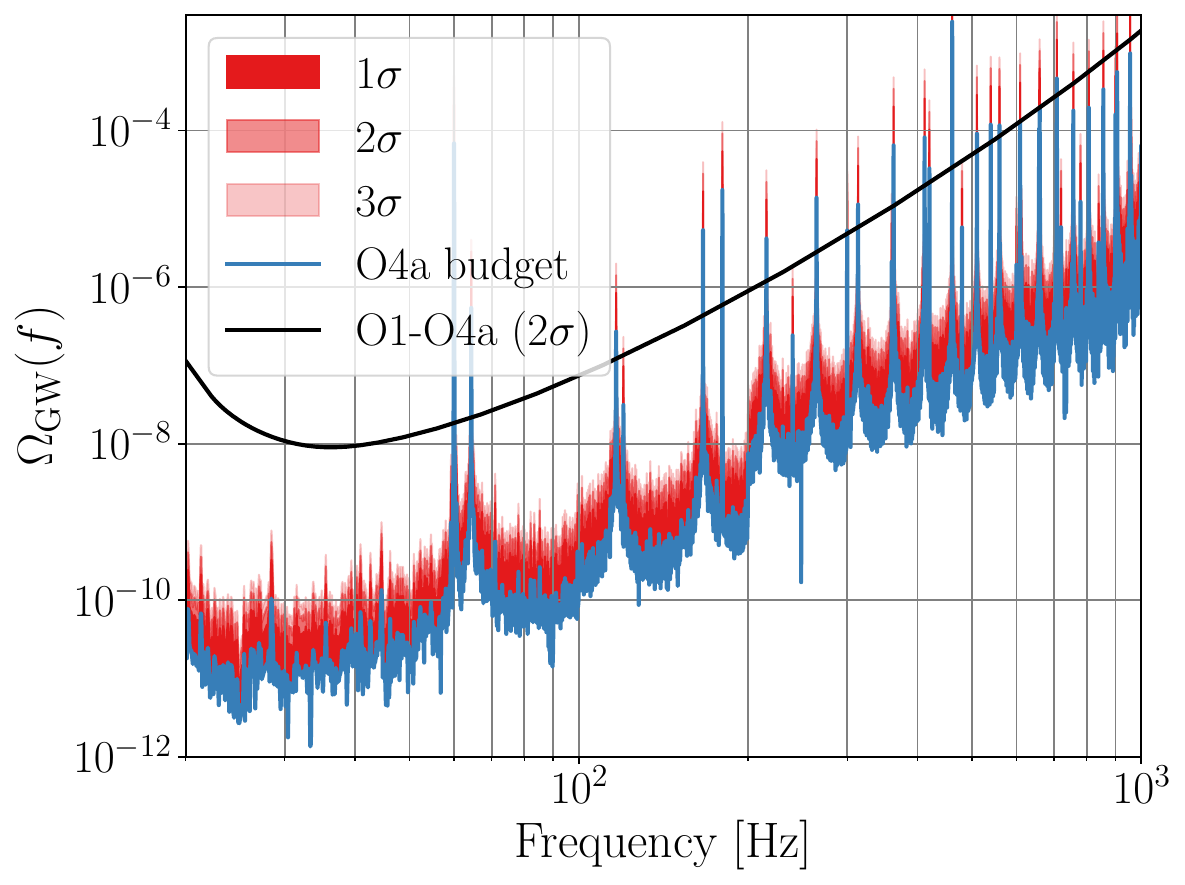}
    \caption{The computed magnetic budget is shown in blue, including $\hat{\Omega}_{\rm mag} + 1\sigma$, $\hat{\Omega}_{\rm mag} + 2\sigma$, and $\hat{\Omega}_{\rm mag} + 3\sigma$ uncertainties (in progressively lighter shades of red). The magnetic budget remains consistently below the $2\sigma$ power-law integrated sensitivity curve (black), except for narrowband features numerically induced by the overlap reduction function (ORF) and harmonics of the 60 Hz power lines. }
    \label{fig:mag-budget}
\end{figure}

The $T_{\rm ITS}$ functions are estimated by each detector site team by injecting far-field magnetic fields within the detector building and measuring the response in the strain channel~\cite{Nguyen_2021, PhysRevD.107.022004, ENV_noise}. 
A value is recorded as a measurement when both the magnetometer sensor and strain channel amplitude spectral densities (ASDs) exceed their respective set thresholds; otherwise, if only the magnetometer ASD exceeds the threshold, the value is treated as an upper limit.
The coupling function is interpolated to the frequencies of the computed magnetic CSD, $\operatorname{Re}\left[\tilde{m}_I^{\star}(f) \tilde{m}_J(f)\right]$. %
This does not create any artifacts in the coupling functions as these have a finer frequency resolution than the CSD. %
We include a calibration factor in the magnetic CSDs\footnote{The calibration factor is an intrinsic conversion factor between the measured current in the magnetometers and the actual magnetic field.} and take the quadrature sum of the magnetic CSDs of four baselines, one for every combination of the two outside magnetometers at each detector used for computing the budget, as a most conservative estimate. %
These quantities are then combined in Eq.~\eqref{eq:mag_coup} to produce the magnetic budget shown in Fig.~\ref{fig:mag-budget}. %
Here the red bands show the uncertainties on the budget, compared to the $2\sigma$ power-law integrated sensitivity (PI) curve~\cite{thrane_sensitivity_2013}, showing the sensitivity of our search to power-law backgrounds. %

The final budget indicates that no correlated (magnetic) noise observation is expected.
Some narrowband features are expected, such as the harmonics of the US power lines at 60 Hz, and the ORF zero-crossings (related to the fact that a single baseline was considered for this analysis). %

While the magnetic budget computation in O3 was performed differently\footnote{In O3, weekly magnetic couplings were used and, due to fluctuations, computing a budget every single week produces a wide uncertainty band instead of a single line~\cite{Nguyen_2021}.}, we can compare the two budgets at an order-of-magnitude level.
At lower frequencies $f < 60$ Hz, the O4a budget is consistently within the band of the O3 budget, but in a narrow frequency band $60\, \mathrm{Hz} < f < 65$ Hz, the O4a budget becomes slightly larger than the maximum value in O3. %
This is related to a zero-crossing of the ORF virtually enhancing the magnetic budget due to its presence in the denominator of Eq.~\eqref{eq:mag_coup}, and does not accurately represent a magnetic response of the detector and thus the actual magnetic budget in O4a.
At higher frequencies $f > 65$ Hz, the O4a budget is broadly consistent with the O3 magnetic budget.

\subsection{Calibration uncertainties}
\label{sec:calibration:uncertainties}
Calibration uncertainties account for both statistical uncertainties and systematic errors arising from the calibration process of LVK data\footnote{Calibration is the procedure used to convert the digital output of a gravitational-wave detector into the relative displacement of the test masses within the detector.}~\cite{Sun:2020wke, Sun:2021qcg}. %
These are incorporated in parameter estimation~\cite{PhysRevD.103.063016,Renzini:2023qtj} by introducing a Gaussian-distributed, positive-definite calibration factor $\Lambda$. As a result, the likelihood presented in Eq.~\eqref{eq: SGWB _likelihood} is modified as (see also App. B of~\cite{Renzini:2023qtj})
\begin{equation}
p(\hOm_0|\Theta, \Lambda)
\propto
\exp
\left[
  -\frac{1}{2}\sum_f \frac{(\hOm_0(f)-\Lambda \OGW(f;\Theta))^2}{\sigma_{0}^2(f)}
\right] \,,
\label{eq:likelihood_calib_factor_1_baseline}
\end{equation}
where $\Lambda$ is marginalized over, assuming the prior %
\begin{equation}
  p(\Lambda) = \frac{1}{\sqrt{2\pi\sigma_\Lambda^2}} \exp\left[-\frac{(\Lambda-1)^2}{2\sigma_\Lambda^2} \right]\,.
\end{equation} 
The calibration variance $\sigma_\Lambda$ is estimated in collaboration with the LIGO Calibration Group. %
Hourly calibration uncertainties provided by the Calibration Group are combined to obtain uncertainties in both the magnitude and phase of the strain at the 68\% confidence level (CL). These uncertainties are frequency-dependent; however, we adopt a conservative approach and select the largest uncertainty across the entire frequency range for both magnitude and phase. The corresponding maximum uncertainties are summarized in Table~\ref{tab:calib_uncertainties}.

\begin{table}[h!]
    \centering
    \begin{tabular}{c|c|c}
        ~ \textbf{Interferometer} ~ &~ \textbf{LHO}~ & ~\textbf{LLO}~  \\ \hline 
        \hline
        Magnitude & ~$6.9\%$~ & ~$4.1\%$~ \\ \hline
        Phase & ~$4.3^\circ$~ & ~$5.1^\circ$ ~ \\ 
    \end{tabular}
    \caption{Maximum calibration uncertainties on the magnitude and phase of detector strain at 68\% confidence level (CL), for the LIGO Hanford (LHO) and LIGO Livingston (LLO) detectors.} 
    \label{tab:calib_uncertainties}
\end{table}

Provided it remains approximately below 5 degrees, the phase uncertainty does not introduce a significant bias in stochastic background estimates~\cite{Yousuf:2023nmz}. Hence we restrict the calculation of calibration uncertainty to the magnitude term only, taking the quadrature sum of the magnitude uncertainties for each detector strain. This yields a final value of $\sigma_\Lambda = 8.1\%$.

\section{Results} 
\label{sec:results}
We present results obtained with the analysis methods described above using data from the O4a observing run combined with data from the O1$-$O3 observing runs. %
We do not find evidence of a background signal in our analyses, as outlined below, hence we set upper limits on several models. %
We perform three different analyses: (i) we calculate the point estimate spectrum and variance of the GWB signal using cross-correlation optimal filtering (Sec.~\ref{sec:results:crossCorrEstimates}) and set upper limits (Sec.~\ref{sec:results:upperLimits}) considering a power-law spectral model for the overall background, both in the case of a fixed known spectral index $\alpha$, and marginalizing over $\alpha$; (ii) we set upper limits on SVT model parameters (Sec.~\ref{sec:results:nonGR}); and (iii) we perform joint inference in the case where a correlated magnetic signal is present in the data together with a GWB (Sec.~\ref{sec:results:jointMagGWB}).

\subsection{Cross-correlation estimates}
\label{sec:results:crossCorrEstimates}
\begin{figure*}[t]
    \centering 
    \includegraphics[width=0.95\textwidth]{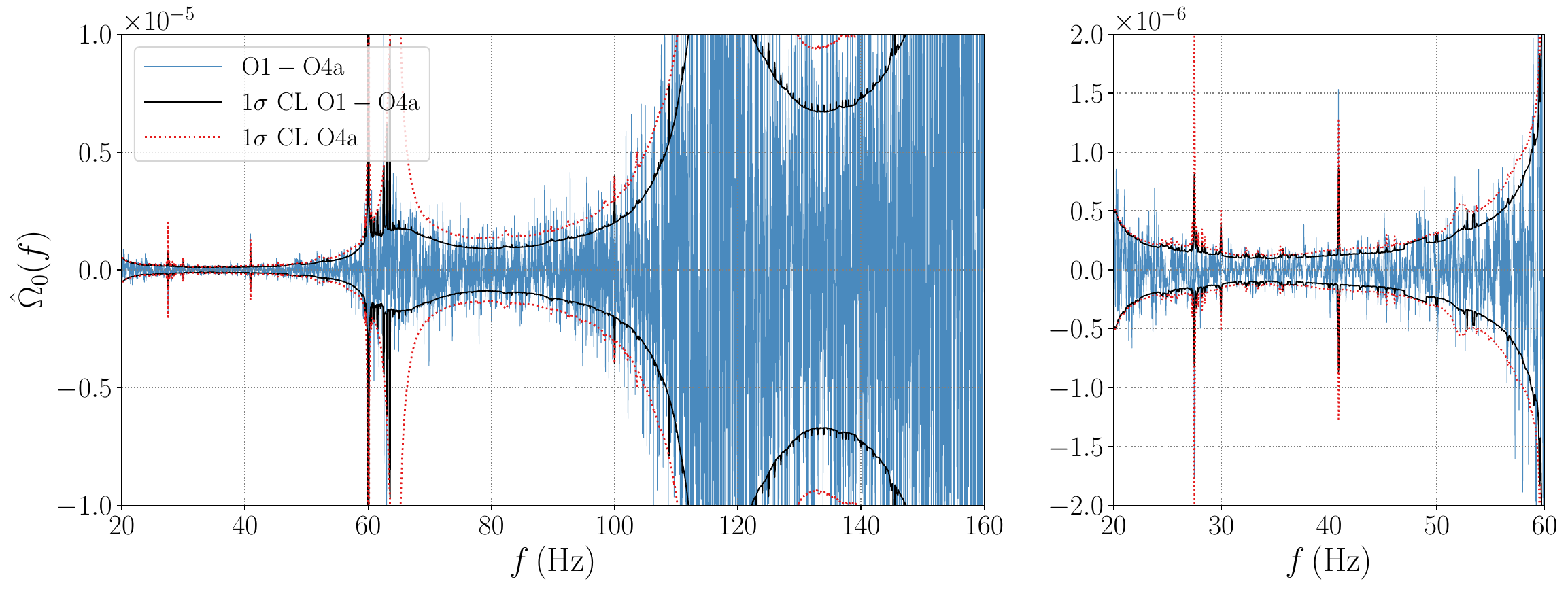} 
    \caption{Cross-correlation spectrum obtained from O1 to O4a data. The black lines denote the boundaries of the $1\sigma$ region: the dotted line shows results calculated from O4a data, while the solid line corresponds to the full O1$-$O4a dataset. The spectrum remains consistent with 0 across all frequency bins, indicating no evidence of a signal. The left panel shows the spectrum over the 20$-$160 Hz range, while the right panel zooms in on frequencies below $60$ Hz to highlight the contribution from O4a data at these frequencies.} 
    \label{fig:cross_corr_spec} 
\end{figure*}

\begin{table*}
    \centering
    \begin{tabular}{c|c|c|c|c}
        ~Power law~ & ~ $\rm f_{99\%} ^ {O4a} [Hz]$ ~ & ~ $\hat{\Omega}^{\rm  O4a}/10^{-9}$ ~ & ~ $\rm f_{99\%} ^ {O1-O4a}$ [Hz] ~ & ~ $ \hat{\Omega}^{\rm O1-O4a}/10^{-9}$~ \\
        \hline
        \hline
        0 & 54.8 & $\ptestOfouraAlzero\pm \sigOfouraAlzero$ & 58.2 & $\ptestOoneOfouraAlzero\pm\sigOoneOfouraAlzero$ \\
        2/3 & 83.6 & $\ptestOfouraAltwoth \pm \sigOfouraAltwoth $ & 86.8 & $\ptestOoneOfouraAltwoth \pm \sigOoneOfouraAltwoth$ \\
        3 & 376.7 & $\ptestOfouraAlthree \pm \sigOfouraAlthree$ & 336.6 & $\ptestOoneOfouraAlthree \pm \sigOoneOfouraAlthree$ \\
    \end{tabular}
    \caption{Search results for an isotropic GWB using cross-correlation optimal filtering for power-law GWBs with $\alpha = {0, 2/3, 3}$. For each case, we report the point estimate and associated $1\sigma$ uncertainty of the amplitude $\hat{\Omega}_{\alpha}$ at $25\,{\rm Hz}$, along with the frequency range from $20\,{\rm Hz}$ to $f_{99\%}$, which contains 99\% of the total sensitivity. These estimates are derived by combining the individual values from independent frequency bins following Eq.~\eqref{eq:broadband_est}. The final two columns show results obtained by combining data from all observing runs (O1$-$O4a) following the prescription shown in~\cite{PhysRevD.104.022004} ( Eq. (7)). Note that Virgo data contributed to the O3 measurements, implying three separate baseline contributions to the O3 results. }
    \label{tab:Frequentist_results}
\end{table*}

We produce broadband estimates $\hOm_{\alpha}$ and associated variance using an optimal filter for three different power-law models motivated by different potential GWB sources:

\begin{itemize}
    \item $\alpha = 0$ corresponds to a scale-invariant signal, which approximately describes a GWB arising from a cosmic string network~\cite{Kibble:1976sj, Sarangi:2002yt, Damour:2004kw, PhysRevLett.98.111101, PhysRevD.97.102002} or slow-roll inflation~\cite{Starobinsky:1979ty, PhysRevD.55.R435, PhysRevD.55.R435, PhysRevD.50.1157} in the LVK frequency band.
    \item $\alpha = 2/3$ characterizes the CBC GWB when the inspiral phase dominates, which is a good approximation in the LVK frequency band~\cite{regimbau:hal-01084645}. However, this approximation may break down for mergers of binaries originating from Population III stars~\cite{Perigois:2020ymr} or for heavy BBH mergers with masses exceeding the pair-instability mass gap~\cite{Ezquiaga:2020tns}. 
    \item $\alpha=3$ is a fiducial choice which provides an approximate description of a subset of astrophysical sources, such as supernovae~\cite{Marassi:2009ib, PhysRevD.73.104024}. It also corresponds to a GWB with a flat strain power spectral density, following $S_h(f) \propto f^{-3} \OGW(f)$~\cite{PhysRevD.59.102001}. This is the spectral shape detectors are most sensitive to by construction, as the optimal filter depends solely on the noise PSD, hence frequencies are weighted according to the detector sensitivity thereby maximizing the cross-correlation search sensitivity. 
\end{itemize}

The point estimate spectrum $\hat{\Omega}_0(f)$ computed as defined in Eq.~\eqref{eq:hOmGW} is shown in Fig.~\ref{fig:cross_corr_spec}. %
The 1$\sigma$ contours calculated from O4a data as well as the entire O1$-$O4a dataset are compatible with 0, as expected in the presence of uncorrelated noise and in the absence of a correlated gravitational-wave signal. %
This behavior is consistent with the expectation that, after applying the data quality cuts described in Section~\ref{sec:dq}, the data remain dominated by Gaussian noise. This is further demonstrated by the Kolmogorov-Smirnov (KS) test, which returns a p-value of $p=0.99$ in favour of Gaussianity. %

Point estimates $\hOm_{\alpha}$ and $1\sigma$ uncertainty using the O4a data only as well as combined results from all the advanced observing runs are presented in Table~\ref{tab:Frequentist_results}. %
Point estimates for all spectral models are consistent with 0 within $1\sigma$, which implies no detection of a GWB. %
We also report the upper edge of the frequency band that contributes 99\% of the sensitivity to each spectral-weighted search, $f_{99\%}$. %

We observe a decrease in variance with increasing spectral index, which is expected since, as mentioned above, our search is most sensitive to a flat strain spectrum ($\alpha = 3$).
Notably, the sensitivity improvement compared to O1$-$O3 results is a factor of $1.6$ for $\alpha = 0$, $1.6$ for $\alpha = 2/3$, and $1.4$ for $\alpha = 3$. %
This is driven by the fact that the most significant sensitivity enhancement in our detectors occurred at lower frequencies~\cite{Capote:2024rmo}, below 40 Hz, as discussed below.

\begin{figure}[t] 
    \centering 
    \includegraphics[width=0.95\linewidth]{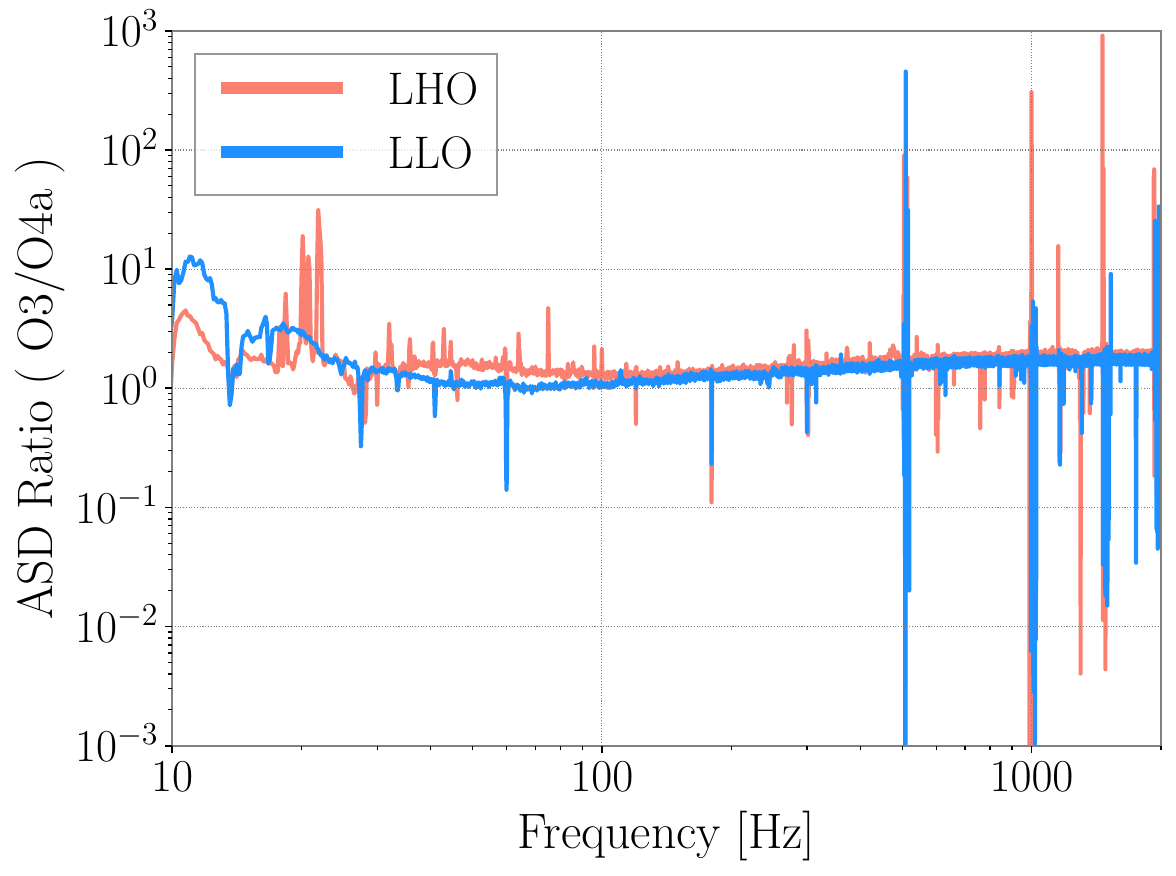} 
    \caption{Ratio of the detector strain sensitivity curves between O3 and O4a for LHO (blue) and LLO (red). A ratio above unity indicates an improvement in sensitivity compared to O3.} 
    \label{fig:sens_curves_ratio} 
\end{figure}

Altogether, these observations indicate that there is no evidence for a detectable GWB in the analyzed data. %
We also note that O4a contributes primarily at frequencies below 60 Hz, as highlighted in the right-hand-side panel of Fig.~\ref{fig:cross_corr_spec} which shows the portion of the spectrum in the $20-60\,{\rm Hz}$ band, where the $1\sigma$ contours from O4a and O1$-$O4a data are almost overlapping. %
The frequency dependence of the sensitivity change between O4a and O3 is described in Fig.~\ref{fig:sens_curves_ratio}, which shows the ratio of the PSDs calculated in the two runs. %
We observe that this ratio lies consistently above 1 in the range $10-60\,\rm Hz$, which matches the range highlighted in Fig.~\ref{fig:sens_curves_ratio} where the O4a data contributes the most to the measurement of a stochastic spectrum. %
The effect of this frequency-dependent improvement in sensitivity to the GWB between O4a and past runs is further quantified below.

\subsection{Upper Limits on the GWB}
\label{sec:results:upperLimits}
Given the absence of evidence for a signal (see Table~\ref{tab:Frequentist_results}), we set upper limits on the amplitude of the GWB, assuming a power-law model as in Eq.~\eqref{eq:plaw_model}.  %
The narrowband estimators $\hOm_0(f)$ and $\sigma_0(f)$  are employed in inference as described in Sec.~\ref{sec:methods:estimator},  where the frequency lines described in Sec.~\ref{sec: notch_list} are notched from the spectra. %
 We explore different priors for both the amplitude and the spectral index. %
Specifically, we consider uniform and log-uniform priors for the amplitude, and uniform and Gaussian priors on the spectral index. %
A log-uniform prior is more sensitive to signals of small amplitude, yielding tighter constraints, while a uniform prior provides more conservative limits. %
In both cases, we set the prior range for the $\Omega_{\rm ref}$ parameter between $\Omega_{\rm ref, min} = 10^{-13}$ and $\Omega_{\rm ref, max} = 10^{-6}$.  

We perform a fixed model analysis, fixing the spectral index to $\{0, 2/3, 3\}$, where each value corresponds to a specific source as indicated above. %
We also perform joint inference on the amplitude and spectral index, where we test both a Gaussian prior and uniform prior on the latter. %
The uniform prior range is set to $\alpha \in [-7, 7]$~\footnote{This prior has been widened with respect to past analyses~\cite{PhysRevD.104.022004}, which employed $\alpha \in [-5, 5]$, to investigate a small mode in the posterior in $\alpha$ which appeared at the edge of the prior range, as discussed below.}. %
In the case of a Gaussian prior on $\alpha$, we impose zero mean and a standard deviation of 
\begin{equation}
    \frac{\log_{10}\Omega_{\rm ref, max} - \log_{10}\Omega_{\rm ref, min}}{2} = 3.5.
\end{equation}

The choice of the prior over $\alpha$ can be understood as follows. %
The log-uniform prior over $\Omega_{\rm ref}$ induces some implicit prior over $\alpha$ that can be shown~\cite{Callister:2017ocg} to be a triangular prior centered on $\alpha = 0$ and non-zero for $ |\alpha| \leq ( \log_{10}\Omega_{\rm ref, max} - \log_{10}\Omega_{\rm ref, min} )$. To avoid a vanishing prior outside of this range, we choose a Gaussian prior for $\alpha$ with a
standard deviation comparable with the triangular prior~\cite{Callister:2017ocg}. %

We set spectral-independent constraints on the signal amplitude by marginalizing over the spectral index posteriors obtained with the Gaussian prior. We take the results obtained with the log-uniform prior on the amplitude as our main result. %

\begin{table*}[t]
    \centering
    \begin{tabular}{c|c|c|c|c|c|c}
         & \multicolumn{3}{|c|}{\textbf{Uniform prior}}& \multicolumn{3}{|c}{\textbf{Log-uniform prior}}  \\ \hline
        $\alpha$ & O1-O4a & O1-O3 & Improvement & O1-O4a & O1-O3 & Improvement \\ \hline \hline
                0 & $\OmrefOoneOfouraULalzeroU \times 10^{-9}$ & $1.7 \times 10^{-8}$ & 2.0 & $\OmrefOoneOfouraULalzerolU \times 10^{-9}$ & $5.8 \times 10^{-9}$ & 2.1  \\ 
        2/3 & $\OmrefOoneOfouraULaltwothU \times 10^{-9}$ & $1.2 \times 10^{-8}$ & 1.9 & $\OmrefOoneOfouraULaltwothlU \times 10^{-9}$ & $3.4 \times 10^{-9}$ & 1.7  \\ 
        3 & $\OmrefOoneOfouraULalthreeU \times 10^{-9}$ & $1.3 \times 10^{-9}$ & 1.3 & $\OmrefOoneOfouraULalthreelU \times 10^{-10}$ & $3.9 \times 10^{-10}$ & 1.2  \\
        Marginalized & $\OmrefOoneOfouraULmargU \times 10^{-8}$ & $2.7 \times 10^{-8}$  & 1.8 & $\OmrefOoneOfouraULmarglU \times 10^{-9}$ & $6.6 \times 10^{-9}$ & 2.3 \\ 
    \end{tabular}
    \caption{Upper limits at 95\% CL on the amplitude $\Omega_{\rm ref}$ of the GWB assuming uniform (left) and log-uniform (right) prior. We quote the upper limits obtained employing data up to the O4a data run (O1$-$O4a), comparing these with those obtained previously~\cite{PhysRevD.104.022004} (O1$-$O3), at individual values of $\alpha$ (which corresponds to imposing a delta function prior on $\alpha$ at the quoted value), and marginalized over the $\alpha$ posterior (assuming a Gaussian prior on $\alpha$). We observe an overall improvement between a factor 1.2 to 2.3. The improvement in upper limits compared to O3 is more significant for smaller spectral indices. This is because the greatest enhancement in detector sensitivity occurred at low frequencies. Furthermore, the difference between the upper limit for $ \alpha = 0$ and the marginalized case is minimal, due to statistical fluctuations from the MCMC. } 
    \label{tab:upper_limits_Omega_ref}
\end{table*}

Table~\ref{tab:upper_limits_Omega_ref} summarizes results for the fixed spectral index and marginalized analyses. %
We set 95\% confidence level upper limits on $\Omega_{\rm ref}$ at $f_{\rm ref} = 25\,{\rm Hz}$ at individual values of $\alpha$, $\Omega_{\rm ref}^{(\alpha)}$, fixing the $\alpha$ prior to a delta function at the chosen values; we also set upper limits marginalized over the Gaussian $\alpha$ posterior. %
For comparison, we report the corresponding upper limits from O1$-$O3 analyses. %
As expected, log-uniform priors on the background amplitude yield tighter constraints compared to uniform priors. %
Broadband constraints at 25 Hz are tightest for $\alpha=3$ as this corresponds to a flat strain power spectrum, yielding the most broadband detectable signal, while constraints are weakest for a scale-invariant spectrum as this signal corresponds to a strain power spectrum with the most negative spectral index out of the set. %
Furthermore, marginalized results over the spectral index yield overall weaker constraints as these admit higher uncertainty on the signal model. %
Overall, we observe improvements in the upper limits ranging from a factor of 1.2 to 2.3. 

\begin{figure}[t] 
    \centering 
    \includegraphics[width=0.4\textwidth]{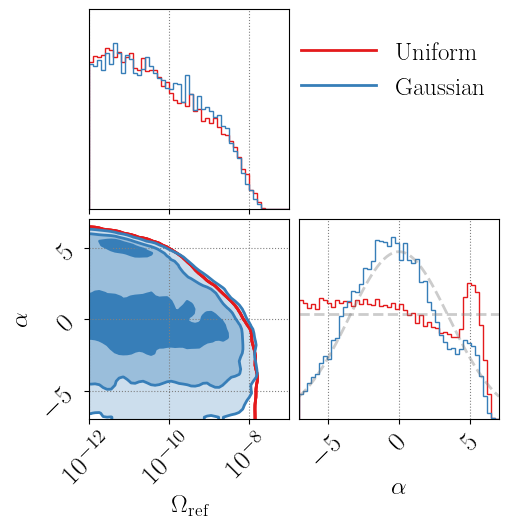} 
    \caption{Posterior distributions on the power-law model parameters $\{\Omega_{\rm ref},\,\alpha\}$ obtained assuming a log-uniform prior on $\Omega_{\rm ref}$ and either a Gaussian (blue) or a uniform (red) prior on $\alpha$. Prior functions for the $\alpha$ parameter are included in the corresponding one-dimensional panel (gray dashed lines).}  
    \label{fig:posteriors_logU_Gaussian} 
\end{figure}
Posterior distributions for $\Omega_{\rm ref}$ and $\alpha$ obtained in these analyses are shown in Fig.~\ref{fig:posteriors_logU_Gaussian} assuming a log-uniform prior on $\Omega_{\rm ref}$ and either a Gaussian or uniform prior on $\alpha$. %
As may be observed in Fig.~\ref{fig:posteriors_logU_Gaussian}, $\alpha$ is not constrained, and the prior on $\alpha$ has little impact on the $\Omega_{\rm ref}$ posteriors. %
The Bayes factor comparing the signal-versus-noise hypothesis to the noise-only hypothesis is $\log_{10}\mathcal{B} = -0.19 \pm  0.03$ ($\log_{10}\mathcal{B} = -0.11 \pm  0.03$) at 68\% confidence level assuming a Gaussian (uniform) prior on $\alpha$, further confirming no evidence of a signal in the data. %

We point out a small narrow mode that appears in the posterior plot at $\alpha=5$, present for both prior assumptions and more evident in the case of a uniform prior on $\alpha$. %
Note that $\alpha=5$ is not associated with any known physical GW signals and usually coincides with the edge of the broad prior range assumed for this analysis, however in this case the prior was broadened to confirm no railing was occurring. %
This mode has been traced back to weak, broad-band residual noise coherence between the LHO and LLO detectors during O4a in the $650-850\,\rm Hz$ band, which corresponds to the frequency range most sensitive to an $\alpha=5$ power-law signal. %
The ORF introduced in Eq.~\eqref{eq:hom_gw} for the LHO--LLO baseline is highly suppressed at these frequencies (to $\sim 0.2-0.1\%$ relative to its magnitude at 1 Hz), strongly suggesting this weak coherence is detector noise and not GW signal. %
Excluding this frequency range from the analysis removes this feature completely. %
The four 60 Hz power-line harmonics in this range, [660, 720, 780, 840] Hz, appear significantly wider and noisier than in past observing runs, hence notches around these lines have been widened accordingly for this analysis. Numerous other weak unidentified noise features such as broadened lines and bumps have been found in both detectors. While these are not loud enough to significantly impact the analysis presented here, as demonstrated by broad-band signal point estimates (Table~\ref{tab:Frequentist_results}) as well as the Bayes' factors above, the significance of this feature will be revisited in the next portion of the O4 data. 

\begin{figure}[t] 
    \centering 
    \includegraphics[width=0.49\textwidth]{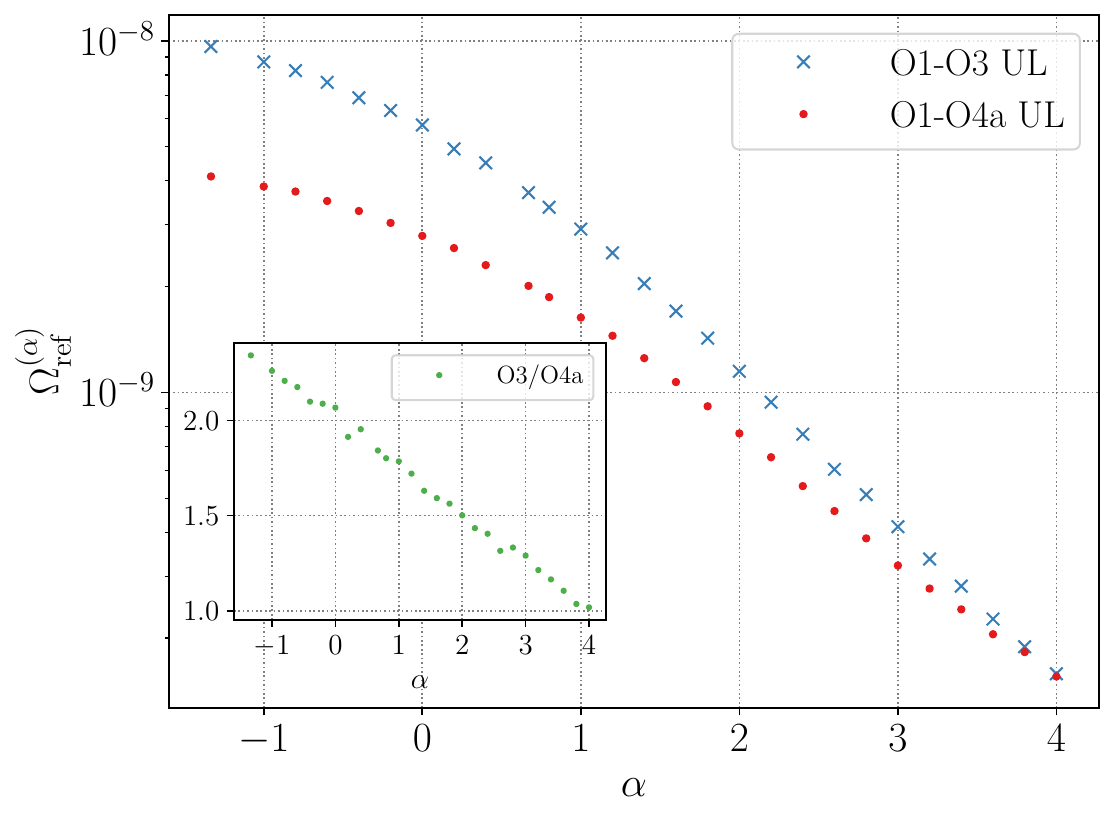} 
    \caption{Upper limits (UL) on $\Omega^{(\alpha)}_{\rm ref}$ at 95\% CL as a function of $\alpha$, obtained assuming the GWB is described by a power-law model using a log-uniform prior on the amplitude $\Omega^{(\alpha)}_{\rm ref}$. The blue curve shows the results using data from O1 to O3, while the red curve includes O4a data as well. The inset plot shows the ratio between the two sets of limits, highlighting improvement in O4a as a function of spectral weighting.} 
    \label{fig:UL_vs_alpha} 
\end{figure}

Finally, Fig.~\ref{fig:UL_vs_alpha} shows upper limits on $\Omega_{\rm ref}$ at 95\% CL as a function of $\alpha$. %
Note that $\alpha$ values in this plot are not directly motivated by specific sources, but rather are chosen to highlight the overall trend: upper limits decrease as the spectral index increases, which is expected due to the extrapolation of the low-frequency sensitivity at higher frequencies, and as a consequence of $\OGW \propto f^3 S_h(f)$. %
Fig.~\ref{fig:UL_vs_alpha} also includes upper limits obtained up to and including the O3 data run. %
The relative improvement between these and the limits obtained including O4a decreases as $\alpha$ increases, indicating that the sensitivity enhancement between runs is more significant at lower frequencies, as discussed above. %

\subsection{Non-GR polarizations}
\label{sec:results:nonGR}
We employ the approach first presented in~\cite{Callister:2017ocg} to constrain potential deviations from GR. The detection of scalar or vector polarization modes would provide direct evidence of a violation of GR, whereas their absence allows us to place stringent constraints on extended theories of gravity. %

Multiple runs with differently polarized backgrounds were conducted. First, we consider a mixed-polarized background, as introduced in Sec.~\ref{sec:methods:pe}. %
In this scenario, the parameters of our model are \begin{equation}
    \Theta = \{\Omega_{\rm ref, T}, \Omega_{\rm ref, V}, \Omega_{\rm ref, S}, \alpha_{\rm T}, \alpha_{\rm V}, \alpha_{\rm S}\} \,.    
\end{equation}
We set a log-uniform prior on the amplitudes $\Omega_{\rm ref, (p)}$, for $\rm (p) = \{T, V, S\}$, over the range $10^{-13}$ to $10^{-6}$ and a Gaussian prior on the spectral indices, centered at 0 with a standard deviation of 3.5. %
We set 95\% CL upper limits on $\Omega_{\rm ref, (p)}$, presented in the third column of Table~\ref{tab:UL_non_GR_pols}. %
We find an improvement in the upper limits of over a factor of 2 with respect to O3 results~\cite{PhysRevD.104.022004}. %
It is worth noting that the upper limit for the scalar-polarized case is less stringent as the ORF for scalar polarization is approximately three times smaller in amplitude than those for other polarizations~\cite{Callister:2017ocg}. %
As a result, our detector network is inherently less sensitive to a scalar-polarized background. 

\begin{table}[h!]
    \centering
    \begin{tabular}{c|c|c|c}

        \textbf{Polarization} & \textbf{O1$-$O3} & \textbf{O1$-$O4a} & \textbf{Improvement} \\ \hline \hline
        Tensor & $6.4\cdot 10^{-9}$ & $2.6\cdot 10^{-9}$ & $\sim 2.5$\\ \hline
        Vector & $7.9\cdot 10^{-9}$ & $2.9\cdot 10^{-9}$ & $\sim2.7$\\ \hline
        Scalar & $2.1\cdot 10^{-8}$ & $7.4\cdot 10^{-9}$ & $\sim2.8$\\ 
    \end{tabular}
    \caption{Upper limits (ULs) at 95\% confidence level (CL) on the amplitude of each individual polarization assuming a mixed-polarized background. The are improved by over a factor of 2 compared to ULs obtained from O1$-$O3 data~\cite{PhysRevD.104.022004}. We observe the UL for the scalar-polarized case is less stringent as the ORF for scalar polarization is smaller in amplitude than those for other polarizations~\cite{Callister:2017ocg}.}
    \label{tab:UL_non_GR_pols}
\end{table}

Additionally, we set upper limits on the amplitudes $\Omega_{\rm ref, T}(25 \rm Hz) \leq 2.9 \cdot 10^{-9}$, $\Omega_{\rm ref, V}(25 \rm Hz) \leq 2.7 \cdot 10^{-9}$ and $\Omega_{\rm ref, S}(25 \rm Hz) \leq 2.9 \cdot 10^{-9}$, assuming single-polarized backgrounds. The upper limits are similar across polarizations due to the reduced detector network in O4a, since only LHO and LLO were operational, unlike O3 which included all three detectors. This limits our ability to distinguish polarization modes.%

The signal versus noise hypothesis Bayes factor yields $\log_{10}\mathcal{B} = -0.87 \pm 0.01$ at 68\% confidence level, indicating no evidence for the presence of a signal in our data. In this case, the signal is defined as the equal-weighted combination of the individual polarization components, $\{\rm T, V, S\}$, as well as their combinations: tensor–vector $(\rm TV)$, tensor–scalar $(\rm TS)$, vector–scalar $(\rm VS)$, and tensor–vector–scalar $(\rm TVS)$. This approach treats all signal configurations equally.

Additionally, the Bayes factor for non-GR versus GR-supported polarized backgrounds yields  $\log_{10}\mathcal{B} = -0.44 \pm 0.03$ at 68\% confidence level. This result suggests that a tensor-polarized background (in accordance with GR) is favoured over vector, scalar, and mixed-polarized alternatives. %
However, in the absence of a detection we expect this result to be driven by the Occam penalty factor introduced by overfitting un-informed parameters in the search. %

\subsection{Joint fit for GWB and magnetic noise}
\label{sec:results:jointMagGWB}
Correlated magnetic noise can provide a significant limitation to the search for an isotropic stochastic signal~\cite{Thrane_2013,Thrane_2014,Coughlin_2018, PhysRevD.107.022004}. %
To mitigate this, the likelihood presented in Eq.~\eqref{eq: SGWB _likelihood} may be extended to jointly model correlated magnetic noise appearing in the GW detectors and a GWB power-law signal, as described in~\cite{PhysRevD.102.102005}:
\begin{equation}
    \Omega_{\rm GW}(f|\Theta) = \Omega_{\rm ref}F_\alpha(f) + \Omega_{\rm MAG}(f|\Theta_{\rm MAG})\,.   
\end{equation}
In Section~\ref{sec:dq:magnetic}, we showed that correlated magnetic noise is not expected to contaminate O4a data. %
Nevertheless, this analysis provides a complementary check by simultaneously fitting both the GWB and the correlated magnetic noise. %
Instead of utilizing the measured coupling functions at the detectors from Section~\ref{sec:dq:magnetic}, we now model the magnetic term $\Omega_{\rm MAG}$ with the parameters $\Theta_{\rm MAG} = \{\kappa_I, \kappa_J, \beta_I, \beta_J\}$, approximating the inside-to-strain magnetic coupling functions $T_{I, \rm ITS}(f)$ as a simple power law,
\begin{equation}
    |T_{I, \rm ITS}(f)| = \kappa_I \left(\frac{f}{10 \mathrm{Hz}}\right)^{-\beta_I}\,,
\end{equation}
where $\kappa_I$ is the magnitude of the amplitude at $10\,\rm Hz$ and $\beta_I$ is the spectral index of the power law.
The prior distribution for $\kappa_I$ is log-uniform in the range $10^{-25}$ to $10^{-22}\,\rm pT^{-1}$ for both detectors and the prior on the spectral index $\beta_I$ is uniform in the range $0-12$ and $1-10$ for HLO and LLO respectively, related to measured minimum and maximum values of the spectral index during magnetic injections. %
As before, the GWB model parameters are $\Omega_{\rm ref}$ and $\alpha$. %
This allows us to estimate either the evidence for a simultaneous measurement of a background and correlated magnetic noise, or estimate the evidence for correlated magnetic noise only. %

We set joint constraints on the power-law GWB and the magnetic correlated noise, employing the same priors for the magnetic parameters as in the O3 analysis~\cite{PhysRevD.104.022004}. %
We find the log Bayes factor comparing a magnetic noise model to a (Gaussian) noise-only model is $\log_{10}\mathcal{B}_N^{\rm MAG} = -0.048$, indicating no evidence in favour of the presence of magnetic noise. %
We also find that for a joint model with a power-law background and magnetic noise modeled as described above that $\log_{10}\mathcal{B}_N^{\rm PL+MAG}= -0.306$, showing no preference for such a model compared to a noise-only model. 

\section{Astrophysical Implications}
\label{sec:astro}
Leveraging the upper limits on the GWB presented in Section \ref{sec:results:upperLimits}, we examine the implications of these results for the GWB of astrophysical origin, namely from the coalescing population of compact binaries including BBHs, BNSs, and NSBHs. %
We compare our upper limits to the most recent predictions for the energy density spectrum associated with different classes of CBC sources. In particular for BBHs, following Ref.~\cite{Callister:2020}, we combine BBH observations in the local Universe with stochastic upper limits to constrain their merger rate at high redshifts, and predict the uncertainty in the BBH background using a fully GW data-driven approach.

\subsection{CBC Model}
\label{sec:astro:cbc}

To model the GWB from binary systems we follow in part the approach of LVK isotropic search papers from the O1 to O3 observing runs~\cite{GW150914_stochastic:2016,PhysRevLett.118.121101,PhysRevD.100.061101, PhysRevD.104.022004}, and additionally we introduce advances in our knowledge of the CBC population~\cite{O3a-pop,O3b-pop,GWTC-4-Population}. %
We categorize the population into classes, each labeled by $k$, to differentiate between BBHs, BNSs and NSBHs. Each class is characterized by distinct values of source parameters, such as masses and spins, which we denote as $\theta_k$. The total astrophysical background is then the sum of contributions from all classes. The contribution of class $k$ to the background can be expressed as an integral over the redshift $z$, given by~\cite{Phinney:2001,Regimbau:2011,Zhu2011,Rosado:2011}
\begin{align}
    \Omega_{\rm GW}(f;\theta_k)=\frac{f}{\rho_c H_0} \int_0^{z_{\rm max}} dz \frac{R_m(z;\theta_k) \langle \frac{dE_{\rm GW}}{df_s}(f_s;\theta_k) \rangle}{(1+z) E(\Omega_\mathrm{m,0},\Omega_{\Lambda,0},z)}, 
    \label{eq:Omega-f}
\end{align}
where $R_m(z;\theta_k)$ denotes the binary merger rate in the source frame per unit comoving volume, $\langle dE_{\rm GW}/df_s(f_s,\theta_k) \rangle$ is the energy spectrum emitted by a single binary evaluated in terms of the source frequency $f_s=(1+z)f$ and averaged over the ensemble properties of source class $k$, and $E(\Omega_\mathrm{m,0},\Omega_{\Lambda,0},z) = \sqrt{\Omega_\mathrm{m,0} (1+z)^3+\Omega_{\Lambda,0}}$ accounts for the dependence of comoving volume on cosmology, assuming $\Omega_{\Lambda,0} = 1-\Omega_\mathrm{m,0}$. We impose an upper redshift cutoff at $z_{\rm max}=10$, beyond which the compact binary merger rate in conventional formation scenarios is expected to be negligible~\cite{Zhu:2011}. We do not consider potential contributions to the GWB from formation channels involving Population III stars~\cite{Martinovic:2021} or primordial black holes~\cite{Mandic:2016,Mukherjee:2021ags}.

The energy spectrum $\langle dE_{\rm GW}/df_s \rangle$ is determined from the strain waveform of the binary system. For computing the backgrounds of BBH and NSBH we use \texttt{IMRPhenomXP}, a phenomenological inspiral-merger-ringdown frequency-domain model for gravitational-wave signals emitted by quasi-circular precessing BBHs~\cite{PhenomX-Pratten:2020,PhemomXP-Pratten:2020}. For BNS background computation, we employ the \texttt{IMRPhenomXP\_NRTidalv3} waveform model, which tapers the merger-ringdown part from the baseline \texttt{IMRPhenomXP} model and has the option to include tidal effects~\cite{Abac:2023,Colleoni:2023,Dietrich:2019}. 
To compute the GWB from CBCs throughout the Universe according to Eq.~(\ref{eq:Omega-f}), we use methods developed for the GWB calculation in~\cite{PhysRevD.104.022004,Ebersold:2025}, as well as \texttt{popstock}, a Python-based module that predicts the background spectrum and its uncertainty given a CBC population and merger rate evolution~\cite{Renzini:2024}.

\subsection{Binary black holes}
\label{sec:astro:cbc:bbh}

To model the BBH background we generally follow the approach in~\cite{PhysRevD.104.022004} with some adjustments in the presumed mass, redshift, and spin distributions of BBHs.

In~\cite{PhysRevD.104.022004} it was assumed that the BBH primary mass distribution is approximated by a mixture between a Gaussian peak near $35\,M_\odot$ and a power-law continuum, that mass ratios are power-law distributed, and, for simplicity, that spins are negligible.
We update these assumptions to match the latest understanding of the black hole population in GWTC-4~\cite{GWTC-4-Results}.
Following~\cite{GWTC-4-Population}, we assume that black hole primary masses follow a mixture between two Gaussians, centered near $10\,M_\odot$ and $35\,M_\odot$, and a broken power-law continuum extending up to $200\,M_\odot$.
Whereas we previously neglected spins, we now include a realistic model for the black hole spin distribution; we follow~\cite{GWTC-4-Population} and assume spin magnitudes are described by a truncated normal distribution and spin orientations by a mixture between isotropic and preferentially-oriented subpopulations.
Black hole mass ratios are again assumed to be power-law distributed.
As in~\cite{PhysRevD.104.022004}, we do not fix the parameters of these models, but instead show a projection for $\Omega_{\rm BBH}(f)$ that self-consistently includes observational uncertainties in the BBH mass, spin, and mass ratio distributions.

Previous forecasts of the BBH background included uncertainties in the ensemble of intrinsic black hole properties and the local BBH merger rate, while imposing a \emph{fixed} fiducial model describing how this rate evolved with redshift.
In~\cite{PhysRevD.104.022004}, for example, we assumed a BBH birth rate proportional to low-metallicity star formation, with a power-law time delay distribution between binary formation and merger.
Here, we revise our projections to additionally reflect systematic uncertainties in the BBH merger rate.
We assume the BBH merger rate density follows the broken power law~\cite{Fishbach2018}
    \begin{equation}
    R_{\rm BBH}(z) = C(\alpha_z, \beta_z, z_{\rm p})
        \frac{ R_0 (1+z)^{\alpha_z}}{1 + \left(\frac{1+z}{1+z_{\rm p}}\right)^{\alpha_z+\beta_z}},
    \label{eq:r-z}
    \end{equation}
where parameters $\alpha_z$ and $\beta_z$ represent the low and high redshift power laws to which the merger rate history asymptotes, respectively, while $z_{\rm p}$ is the redshift at which the merger rate history peaks. %
Specifically, Eq.~\eqref{eq:r-z} implies that $R_{\rm BBH}(z) \propto (1+z)^{\alpha_z}$ for redshifts $z\ll z_{\rm p}$ and $R_{\rm BBH}(z) \propto (1+z)^{-\beta_z}$ for $z\gg z_{\rm p}$, and the normalization constant $C(\alpha_z, \beta_z, z_{\rm p})$ is defined such that $R_{\rm BBH}(z) = R_0$ at $z=0$. %
This parametrization allows us to cover a broad range of theoretical predictions for the redshift evolution of the BBH merger rate.
It accommodates scenarios involving isolated binary evolution, where the merger rate closely follows the metallicity-dependent cosmic star formation rate with a distribution of time delays between formation and merger~\cite{Dominik:2013,Mapelli:2017,Santoliquido:2020}.
At the same time, it can represent dynamically assembled mergers in dense star clusters, which are less directly tied to star formation and often exhibit a merger rate that peaks at higher redshifts~\cite{Mapelli:2021,Kou:2024}. 

\begin{figure*}[t] 
    \centering 
    \includegraphics[width=0.95\textwidth]{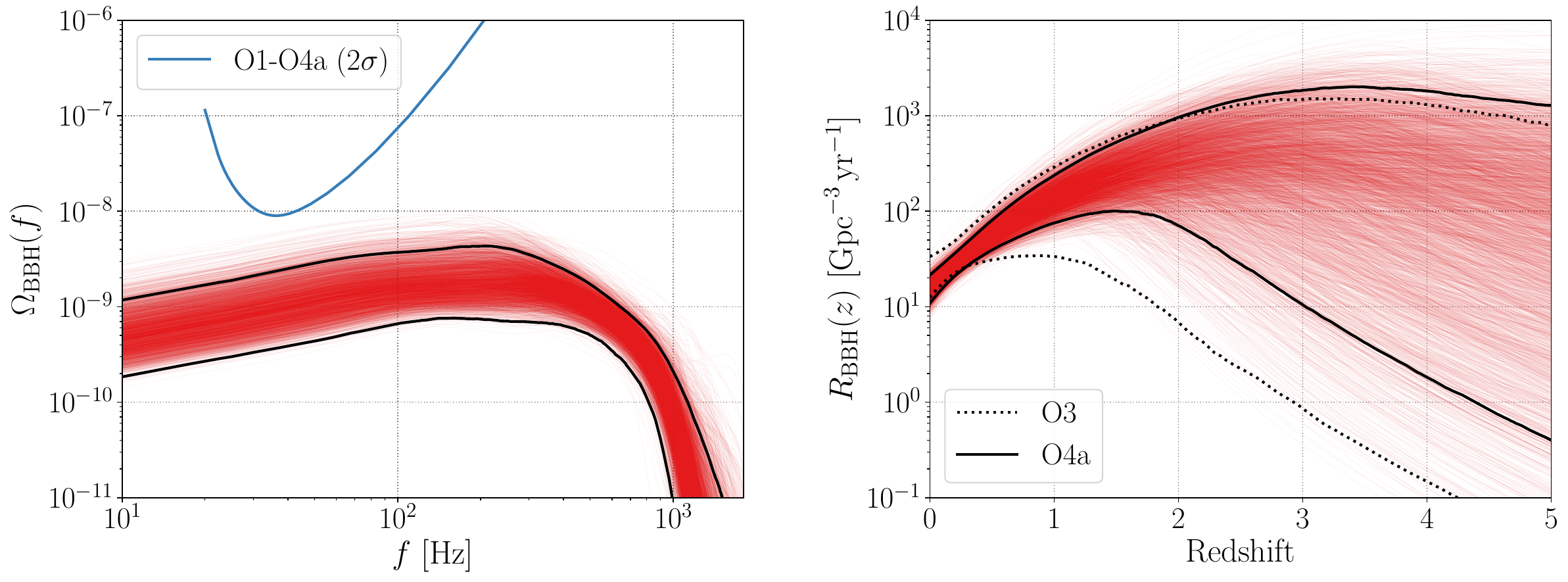}\caption{
    \textit{Left:}
    Forecast for the stochastic energy-density spectrum of the BBH population, given the population of direct BBH detections within the LVK GWTC-4 catalog~\cite{GWTC-4-Population} and the GWB upper limit presented here.
    The black curves indicate 90\% credible bounds on $\Omega_\mathrm{BBH}(f)$, while red traces show individual posterior samples on the energy-density spectrum, obtained by sampling over the possible mass, spin, and redshift distributions of BBH mergers.
    For comparison, the blue curve indicates the $2\sigma$ PI curve calculated using O1$-$O4a HLV data~\cite{thrane_sensitivity_2013}, integrated over all observing runs to date.
    \textit{Right:}
    Inferred redshift evolution of the BBH merger rate density, measured hierarchically using the direct BBH detections in GWTC-4 and the GWB upper limit.
    Solid black and red lines again mark 90\% credible bounds and individual posterior samples, respectively, while dotted black lines denote the 90\% credible bounds obtained previously with GWTC-3~\cite{O3b-pop}.
    } 
    \label{fig:bbh-forecast} 
\end{figure*}

\begin{figure*}[t] 
    \centering 
    \includegraphics[width=0.95\textwidth]{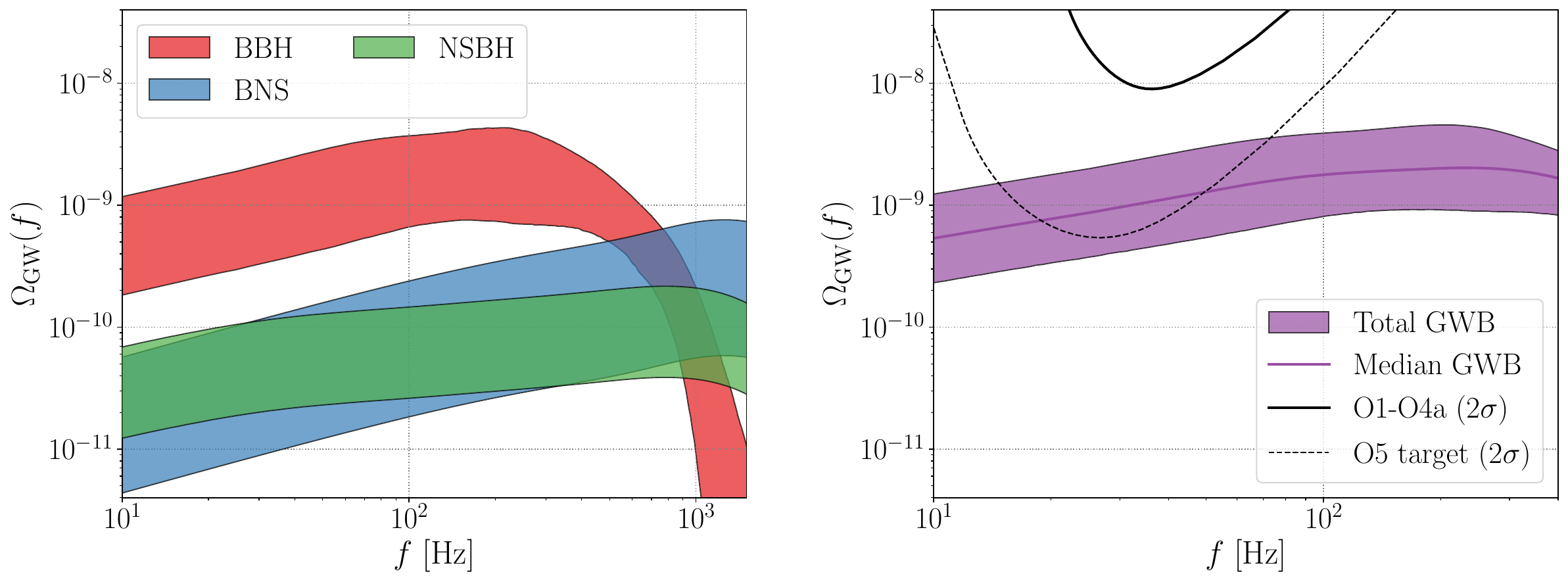}
    \caption{
    \textit{Left:}
    Projected astrophysical GWB contributions from BBH, BNS and NSBH coalescences. Shaded bands represent the 90\% credible bounds for each CBC class.
    \textit{Right:}
    The solid purple line shows the median estimate of $\Omega_\mathrm{BBH+BNS+NSBH} (f)$ as a function of frequency, while the shaded purple band illustrates 90\% credible uncertainties. We also show the $2\sigma$ PI sensitivity curve calculated using data from stochastic isotropic searches O1 through O4a, considering the three HLV baselines. We include the target $2\sigma$ PI curve for O5, assuming the target sensitivity curves for the HLV baselines as detailed in~\cite{t2000012} and~\cite{O5VirgoNoiseASDCurve} .
    } 
    \label{fig:bns_nsbh_forecast}
\end{figure*}

We combine all direct detections of BBHs in GWTC-4 with our updated constraints on the astrophysical GWB to measure the underlying distributions of BBH primary masses, mass ratios, spins, and redshifts.
We assume these distributions are well-described by the models listed above, and we follow the methods presented in~\cite{GWTC-4-Population,Callister:2020, Turbang:2023tjk, Lalleman:2025xcs}.
In order to self-consistently combine data from direct BBH detections with constraints on the GWB, we adopt a factorized likelihood given by 
\begin{equation}
    p(\hat{\Omega}_0, \{d_i\}|\Lambda_{\rm BBH}) = p_{\rm BBH}(\{d_i\}|\Lambda_{\rm BBH}) \times p_{\rm stoch}(\hat{\Omega}_0|\Lambda_{\rm BBH})\,.
    \label{eq:implilikelihood}
\end{equation}
Here, $p_{\rm stoch}(\hat{\Omega}_0|\Lambda_{\rm BBH})$ is the likelihood of our cross-correlation measurements [see Eq.~\eqref{eq: SGWB _likelihood}], given a BBH population model denoted $\Lambda_{\rm BBH}$.
Meanwhile, $p_{\rm BBH}(\{d_i\}|\Lambda_{\rm BBH})$ is the likelihood of the direct compact binary detections comprising GWTC-4~\cite{GWTC-4-Population}.
In practice, rather than evaluating Eq.~\eqref{eq:implilikelihood} directly, we sample the likelihood in a two step process, first sampling from the posterior $p_{\Lambda_{\rm BBH}| \rm BBH}(\{d_i\})$ and then rejection sampling these draws using $p_{\rm stoch}(\hat{\Omega}_0|\Lambda_{\rm BBH})$.
The resulting $\Omega_{\rm BBH}(f)$ spectra calculated from each posterior sample $\Lambda$ after rejection sampling are shown in the left-hand side of Fig.~\ref{fig:bbh-forecast}.
At a reference frequency of $25\,\mathrm{Hz}$, we predict $\Omega_\mathrm{BBH}(25\,\mathrm{Hz}) = \bbhMedian^{+\bbhUpperError}_{-\bbhLowerError}\times 10^{\bbhPower}$ before rejecting according to the GWB upper limits. %
After rejection sampling, the predicted BBH GWB amplitude is
$\Omega_\mathrm{BBH}(25\,\mathrm{Hz}) = \bbhfinalMedian^{+\bbhfinalUpperError}_{-\bbhfinalLowerError}\times 10^{\bbhfinalPower}$. %
Here and henceforth, all uncertainties are quoted as 90\% credible intervals.

The inferred BBH merger rate $R_\mathrm{BBH}(z)$ is shown in the right side of Fig.~\ref{fig:bbh-forecast}.
Black and red curves again show 90\% credible bounds and individual posterior samples, respectively.
The present-day merger rate is inferred to be $R_0 = 15.4^{+6.0}_{-4.5} \, \mathrm{Gpc}^{-3}\, \mathrm{yr}^{-1}$.
For comparison, dotted black curves indicate 90\% credible bounds obtained previously when analyzing BBH mergers in GWTC-3~\cite{GWTC3,O3b-pop}.
New results are consistent with previous estimates, although it is now found that the rate with which $R_\mathrm{BBH}(z)$ increases ($\alpha_z = \alphaMedianOFoura^{+\alphaUpperErrorOFoura}_{-\alphaLowerErrorOFoura}$) is near the upper end of previous bounds ($\alpha_z =\alphaMedianOThree^{+\alphaUpperErrorOThree}_{-\alphaLowerErrorOThree}$). %
The high redshift parameters $\beta_z$ and $z_{\rm p}$ remain poorly constrained ($\beta_z=\betaMedianOFoura^{+\betaUpperErrorOFoura}_{-\betaLowerErrorOFoura}$, $z_{\rm p}=\zpeakMedianOFoura^{+\zpeakUpperErrorOFoura}_{-\zpeakLowerErrorOFoura}$).

The forecasted $\Omega_\mathrm{BBH}(25\,\mathrm{Hz})$ reported above is approximately a factor of two larger than previously estimated in~\cite{PhysRevD.104.022004} and~\cite{O3b-pop}.
This increase is due primarily to two effects.
The increased sample of direct compact binary detections indicates that the BBH merger rate likely increases more quickly with redshift than previously measured. %
Secondly, whereas past forecasts used simple parametric estimates for the energy spectrum of a given binary~\cite{Ajith2011}, our updated forecasts adopt the significantly more mature \texttt{IMRPhenomXP} waveform model to compute radiated energy.

The non-detection of a stochastic background provides an additional constraint on the population history of BBHs, effectively ruling out scenarios with high merger rates at large redshifts that would otherwise yield a detectable background signal~\cite{Callister:2020}.
While the background upper limits effectively do not impact the inferred merger rate, these imply a modest tightening of the upper uncertainty bound on the BBH background amplitude as described above.
This result implies that, when modeling the BBH merger history via Eq.~\eqref{eq:r-z}, the direct detections already require $\alpha_z$ to be sufficiently small that most possible astrophysical backgrounds lie below the current detection threshold.
In future observing runs, the exact degree to which stochastic search results are or are not informative will depend on the achieved sensitivity and run duration, as well as updated inference of $\alpha_z$.

\subsection{Binary neutron stars}
\label{sec:astro:cbc:bns}

Since neither the local merger rate nor the mass and spin distribution of BNS mergers are well constrained from current GW catalogs, we cannot adopt detailed data-driven models of the BNS population.
Instead, we forecast the astrophysical background from BNSs using the methodology of~\cite{PhysRevD.104.022004}, which adopts a fixed star formation history convoluted with a time delay distribution and simplified assumptions for mass and spin.
We assume that BNS progenitor formation follows the star formation rate described in~\cite{Madau:2016} (Eq.(\ref{eq:r-z}) with $\alpha_z = 2.6$, $\beta_z = 3.6$ and $z_p = 2.2$), with a distribution $p(t_d) \sim t_d^{-1}$ of time delays $t_d$ between binary formation and merger, where 20 Myr $< t_d <$ 13.5 Gyr.
Following~\cite{GWTC-4-Population}, we assume a simple fixed population model, with a uniform mass distribution between $1-2.5 \, M_\odot$ and isotropic spin orientations with uniformly distributed spin magnitudes below 0.4. Under these assumptions, the local merger rate is estimated to be $\bnsrateMedian^{+\bnsrateUpperError}_{-\bnsrateLowerError}$ $\mathrm{Gpc}^{-3} \mathrm{yr}^{-1}$.
When modeling $\Omega_\mathrm{BNS} (f)$, we do not consider any tidal effects nor contributions from a potential post-merger signal~\cite{Lehoucq:2025}.
These modeling choices lead to an estimated BNS GWB shown in Fig.~\ref{fig:bns_nsbh_forecast}, specifically at the reference frequency of 25 Hz, we predict $\Omega_\mathrm{BNS}(25 \mathrm{Hz}) = {\bnsMedian^{+\bnsUpperError}_{-\bnsLowerError} \times 10^{\bnsPower}} $.

\subsection{Neutron star-black hole binaries}
\label{sec:astro:cbc:nsbh}

Similar to BNSs, few compact binaries with masses consistent with NSBH systems have been observed~\cite{GW190814,LIGO_NSBH:2021,LIGO_GW230529:2024}, and so the details of the NSBH population remain largely unknown.
As in the BNS case above, we again follow the simplified modeling approach undertaken previously in~\cite{O3b-pop}.
The black hole masses are drawn from a log-uniform distribution between 3 and $50 \, M_\odot$, and neutron star masses from a uniform distribution between $1-2.5 \, M_\odot$. Black hole spins are assumed to be isotropically distributed with uniform spin magnitude below 0.99, while the neutron star spin follows the same distribution as for BNS.
The energy emission from NSBH binaries is calculated using the BBH waveform model \texttt{IMRPhenomXP}, which provides a reliable approximation, especially for systems with large mass ratios~\cite{Foucart:2013}. However, the model does not include tidal deformability or the possibility of tidal disruption of the neutron star, effects that can affect the emitted GW signal from binaries with more comparable component masses~\cite{Pannarale:2015}.
For the given population model, the local merger rate of NSBH binaries is estimated to be {$\nsbhrateMedian^{+\nsbhrateUpperError}_{-\nsbhrateLowerError}$} $\mathrm{Gpc}^{-3} \mathrm{yr}^{-1}$~\cite{GWTC-4-Population}. Moreover, we assume that the rate evolves in redshift in a manner identical to BNS systems as described above.

With these assumptions we find $\Omega_\mathrm{NSBH}(25 \mathrm{Hz}) = {\nsbhMedian^{+\nsbhUpperError}_{-\nsbhLowerError} \times 10^{\nsbhPower}}$. The 90\% credible band over the whole frequency range is displayed in Fig.~\ref{fig:bns_nsbh_forecast}.
We note that $\Omega_\mathrm{NSBH}(f)$ is predicted to be significantly flatter than in past estimates, deviating noticeably from a canonical $f^{2/3}$ power law above $\sim 30\,\mathrm{Hz}$.
This change is primarily due to our adoption of an updated waveform model, in which the energy spectra of highly unequal-mass binaries are suppressed at high frequencies.

\subsection{Total compact binary background}

In Fig.~\ref{fig:bns_nsbh_forecast} (right hand side), we present an updated estimate of the combined CBC GWB. Our model predicts this background at $\Omega_\mathrm{BBH+BNS+NSBH}(25 \mathrm{Hz}) = {\cbcMedian^{+\cbcUpperError}_{-\cbcLowerError} \times 10^{\cbcPower}} $. Additionally, we show the estimated sensitivity of the search described in this paper by presenting the $2\sigma$ power-law integrated sensitivity curve including data from O1$-$O4a. Although our estimate for the background amplitude is below current limits, it may become accessible with the A+ configuration of the LIGO detectors. This may be seen from the comparison shown in Fig.~\ref{fig:bns_nsbh_forecast} with the target $2\sigma$ power-law integrated sensitivity curve for the O5 data run (dashed curve), which includes substantial planned updates for the LIGO sites~\cite{Capote:2024rmo,t2000012} and the Virgo detector~\cite{O5VirgoNoiseASDCurve}, and assumes one year of uninterrupted data. %
It is important to note that these are \emph{target} sensitivities that are expected to be reached by the end of the observing run, hence the O5 target sensitivity curve in Fig.~\ref{fig:bns_nsbh_forecast} should be interpreted as an optimistic limit for the observing scenario. 

\label{sec:astro:cbc:totalCBC}

\section{Conclusions}
\label{sec:concl}
In this paper, we presented an isotropic search conducted on data spanning from the first LVK observing run (O1) to the first part of the fourth observing run (O4a), incorporating 115.51 days of new coincident LIGO-Hanford and LIGO-Livingston observation.
No evidence for a GWB is found, and we limit the net energy density of the GWB to $\Omega_{\rm ref}(25\,\mathrm{Hz}) \leq \OmrefOoneOfouraULmarglU \times 10^{-9}$ under a log-uniform prior ($\leq 1.5\times10^{-8}$ under a uniform prior), an improvement of approximately a factor of 2 relative to previous bounds~\cite{PhysRevD.104.022004}. %

We additionally constrain alternative theories of gravity by searching for non-standard gravitational-wave polarizations, including scalar and vector modes beyond general relativity's tensor predictions. Our analysis improves existing bounds on these exotic polarization states by factors of two compared to previous datasets. However, the absence of any detectable background prevents us from distinguishing between standard tensor-only scenarios and models incorporating additional polarization modes.

As GWB searches improve in sensitivity, previously negligible terrestrial noise sources, including magnetic coupling from Schumann resonances and lightning activity, might contaminate our results and require characterization and possible mitigation. %
To quantify the potential impact of magnetic couplings on our result, we combine measurements of the magnetic environments surrounding LHO and LLO with measured transfer functions between external magnetic fields and measured strain, estimating a net magnetic noise budget. %
We find that, given current search sensitivities, contamination due to correlated magnetic environments does not significantly impact our results.

We conclude by examining how close our sensitivity has come to detecting the expected astrophysical background from the population of merging compact binaries across the Universe. %
We present updated estimates of energy-density spectra arising from distant binary black holes, binary neutron stars, and neutron-star black hole binaries, incorporating updated measurements of these sources' merger rates and demographics.
Combining these three source classes, we estimate a total astrophysical background amplitude $\Omega_\mathrm{BBH+BNS+NSBH}(25 \mathrm{Hz}) = {\cbcMedian^{+\cbcUpperError}_{-\cbcLowerError} \times 10^{\cbcPower}}$, consistent with but marginally smaller than previous estimates~\cite{O3b-pop}.
This decrease is primarily driven by lower inferred BNS and NSBH merger rates, given the lack of such systems detected during O4a.
This estimate lies at least a factor of three below the sensitivity of our search, and is thus consistent with a current non-detection of a GWB.

Several companion papers further extend the methodology presented here and elaborate on additional implications of a GWB non-detection.
In this work, we have limited ourselves to exploring a purely isotropic stochastic background.
Alternate analyses targeting anisotropic backgrounds are presented in~\cite{Anisotropic}.
Additionally, although we have commented on the astrophysical background expected from merging compact binaries, there exist a host of predicted mechanisms that may yield GWBs of cosmological origin.
Implications of our updated constraints on the GWB on such cosmological backgrounds are discussed in~\cite{CosmoImplications}.

\section{Acknowledgements}
This material is based upon work supported by NSF's LIGO Laboratory, which is a
major facility fully funded by the National Science Foundation.
The authors also gratefully acknowledge the support of
the Science and Technology Facilities Council (STFC) of the
United Kingdom, the Max-Planck-Society (MPS), and the State of
Niedersachsen/Germany for support of the construction of Advanced LIGO 
and construction and operation of the GEO\,600 detector. 
Additional support for Advanced LIGO was provided by the Australian Research Council.
The authors gratefully acknowledge the Italian Istituto Nazionale di Fisica Nucleare (INFN),  
the French Centre National de la Recherche Scientifique (CNRS) and
the Netherlands Organization for Scientific Research (NWO)
for the construction and operation of the Virgo detector
and the creation and support  of the EGO consortium. 
The authors also gratefully acknowledge research support from these agencies as well as by 
the Council of Scientific and Industrial Research of India, 
the Department of Science and Technology, India,
the Science \& Engineering Research Board (SERB), India,
the Ministry of Human Resource Development, India,
the Spanish Agencia Estatal de Investigaci\'on (AEI),
the Spanish Ministerio de Ciencia, Innovaci\'on y Universidades,
the European Union NextGenerationEU/PRTR (PRTR-C17.I1),
the ICSC - CentroNazionale di Ricerca in High Performance Computing, Big Data
and Quantum Computing, funded by the European Union NextGenerationEU,
the Comunitat Auton\`oma de les Illes Balears through the Conselleria d'Educaci\'o i Universitats,
the Conselleria d'Innovaci\'o, Universitats, Ci\`encia i Societat Digital de la Generalitat Valenciana and
the CERCA Programme Generalitat de Catalunya, Spain,
the Polish National Agency for Academic Exchange,
the National Science Centre of Poland and the European Union - European Regional
Development Fund;
the Foundation for Polish Science (FNP),
the Polish Ministry of Science and Higher Education,
the Swiss National Science Foundation (SNSF),
the Russian Science Foundation,
the European Commission,
the European Social Funds (ESF),
the European Regional Development Funds (ERDF),
the Royal Society, 
the Scottish Funding Council, 
the Scottish Universities Physics Alliance, 
the Hungarian Scientific Research Fund (OTKA),
the French Lyon Institute of Origins (LIO),
the Belgian Fonds de la Recherche Scientifique (FRS-FNRS), 
Actions de Recherche Concert\'ees (ARC) and
Fonds Wetenschappelijk Onderzoek - Vlaanderen (FWO), Belgium,
the Paris \^{I}le-de-France Region, 
the National Research, Development and Innovation Office of Hungary (NKFIH), 
the National Research Foundation of Korea,
the Natural Sciences and Engineering Research Council of Canada (NSERC),
the Canadian Foundation for Innovation (CFI),
the Brazilian Ministry of Science, Technology, and Innovations,
the International Center for Theoretical Physics South American Institute for Fundamental Research (ICTP-SAIFR), 
the Research Grants Council of Hong Kong,
the National Natural Science Foundation of China (NSFC),
the Israel Science Foundation (ISF),
the US-Israel Binational Science Fund (BSF),
the Leverhulme Trust, 
the Research Corporation,
the National Science and Technology Council (NSTC), Taiwan,
the United States Department of Energy,
and
the Kavli Foundation.
The authors gratefully acknowledge the support of the NSF, STFC, INFN and CNRS for provision of computational resources. %
This work was supported by MEXT,
the JSPS Leading-edge Research Infrastructure Program,
JSPS Grant-in-Aid for Specially Promoted Research 26000005,
JSPS Grant-in-Aid for Scientific Research on Innovative Areas 2402: 24103006,
24103005, and 2905: JP17H06358, JP17H06361 and JP17H06364,
JSPS Core-to-Core Program A.\ Advanced Research Networks,
JSPS Grants-in-Aid for Scientific Research (S) 17H06133 and 20H05639,
JSPS Grant-in-Aid for Transformative Research Areas (A) 20A203: JP20H05854,
the joint research program of the Institute for Cosmic Ray Research,
University of Tokyo,
the National Research Foundation (NRF),
the Computing Infrastructure Project of the Global Science experimental Data hub
Center (GSDC) at KISTI,
the Korea Astronomy and Space Science Institute (KASI),
the Ministry of Science and ICT (MSIT) in Korea,
Academia Sinica (AS),
the AS Grid Center (ASGC) and the National Science and Technology Council (NSTC)
in Taiwan under grants including the Science Vanguard Research Program,
the Advanced Technology Center (ATC) of NAOJ,
and the Mechanical Engineering Center of KEK.

\bibliographystyle{apsrev4-1}
\bibliography{References}
\end{document}